\documentclass[twocolumns]{aa}
\usepackage{graphicx}
\usepackage{amsmath}
\usepackage{txfonts}
\usepackage{natbib}
\usepackage{url}
\bibpunct{(}{)}{;}{a}{}{,}

\def\spose#1{\hbox to 0pt{#1\hss}}
\def\lta{\mathrel{\spose{\lower 3pt\hbox{$\mathchar"218$}} \raise 2.0pt\hbox{$\mathchar"13C$}}}
\def\gta{\mathrel{\spose{\lower 3pt\hbox{$\mathchar"218$}} \raise 2.0pt\hbox{$\mathchar"13E$}}}

\begin{document}

\title{The best fit for the observed galaxy Counts-in-Cell distribution function}
\titlerunning{The best fit for the observed galaxy Counts-in-Cell distribution function}

\author{Llu\'{\i}s Hurtado-Gil\inst{\ref{inst1},\ref{inst2},\ref{inst3}}  \and Vicent~J.~Mart\'{\i}nez \inst{\ref{inst2},\ref{inst3}}  \and Pablo~Arnalte-Mur \inst{\ref{inst2},\ref{inst3}} \and   Mar\'{\i}a-Jes\'us Pons-Border\'{\i}a \inst{\ref{inst4}} \and Crist\'obal Pareja-Flores \inst{\ref{inst5}} \and Silvestre Paredes \inst{\ref{inst6}}}
\authorrunning{Llu\'{\i}s Hurtado et al.}

\institute{
Departamento de Matem\'atica Aplicada y Estad\'{\i}stica, Universidad CEU San Pablo, C/  Julian Romea 23, 28003 Madrid, Spain 
\email{lluis.hurtadogil@ceu.es}
\label{inst1}
\and
Observatori Astron\`omic, Universitat de Val\`encia, C/ Catedr\'atico Jos\'e Beltran, 2, 46980 Paterna (Val\`encia), Spain
\label{inst2}
\and
Departament d'Astronomia i Astrof\'{\i}sica, Universitat de Val\`encia, 46100-Burjassot, Val\`encia, Spain \label{inst3}
\and
Departamento de Matem\'atica Aplicada, Facultad de Estudios  Estad\'{\i}sticos, Universidad Complutense de Madrid, Madrid, Spain \label{inst4}
\and
Departamento de Sistemas Inform\'aticos y Computaci\'on, Facultad de Estudios Estad\'isticos, Universidad Complutense de Madrid, Madrid, Spain  \label{inst5}
\and
Departamento de Matem\'atica Aplicada y Estad\'{\i}stica, Universidad Polit\'ecnica de Cartagena, C/ Dr. Fleming s/n, Cartagena, Spain \label{inst6}
}

\date{Received XXX; accepted YYY}

\abstract
{The Sloan Digital Sky Survey (SDSS) is the first dense redshift survey encompassing a volume large enough to find the best analytic probability density function that fits the galaxy Counts-in-Cells distribution  $f_V(N)$, the frequency distribution of galaxy counts in a volume $V$.}
{ Different analytic functions have been previously proposed that can account for some of the observed features of the observed frequency counts, but fail to provide an overall good fit to this important statistical descriptor of the galaxy large-scale distribution. Our goal is to find the probability density function that better fits the observed Counts-in-Cells distribution $f_V(N)$.}
{We have made a systematic study of this function applied to several samples drawn from the SDSS.  We show the effective ways to deal with incompleteness of the sample (masked data) in the calculation of $f_V(N)$. We use LasDamas simulations to estimate the errors in the calculation. We test four different distribution functions to find the best fit: the Gravitational Quasi-Equilibrium distribution, the Negative Binomial Distribution, the Log Normal distribution and the Log Normal Distribution including a bias parameter. In the two latter cases, we apply a shot-noise correction to the distributions assuming the local Poisson model.}
{We show that the best fit for the Counts-in-Cells distribution function is provided by the Negative Binomial distribution. 
 In addition, at large scales the Log Normal distribution modified with the inclusion of the bias term also performs a satisfactory fit of the empirical values of  $f_V(N)$.Our results demonstrate that the inclusion of a bias term in the Log Normal distribution is necessary to fit the observed galaxy Count-in-Cells distribution function.}
{}


   \maketitle
%

\section{Introduction}
\label{sec:intro} 
One of the first statistics applied to study the galaxy clustering in the 
very early galaxy catalogs (that were built on the projected celestial sphere) was the Counts-in-Cells (CiC) method. \cite{1934ApJ....79....8H} was the first to notice that the distribution of galaxy counts in two-dimensional cells could be well approximated by a lognormal distribution. This technique permits to describe the spatial distribution of galaxies in a way that is complementary to other descriptors of the galaxy clustering such as the correlation function or the power spectrum \citep{1979MNRAS.189..831W,1980lssu.book.....P}. The count probability distribution function $f_V(N)$ gives the probability that a randomly placed volume in the universe will contain exactly $N$ galaxies. For $N=0$, this function is known as the void-probability function \citep{1987ApJ...320...13M} and it is of particular interest, since is related with higher order correlation functions \citep{1979MNRAS.189..831W} and provides a simple approach to establish hierarchical scaling relations \citep{1989A&A...220....1B,2004MNRAS.352..828C}. \cite{1994ApJ...425....1F} have shown that correlation functions can be measured from the moments of $f_V(N)$. Several distribution functions have been used to model the Counts-in-Cells galaxy distribution \citep{2011ApJ...729..123Y,2016A&A...588A..51B} and the Counts-in-Cells matter distribution \citep{2017MNRAS.466.1444C}, or used this distribution to estimate galaxy-dark matter bias \citep{2000ApJ...540...62S,2005A&A...442..801M,2011ApJ...731..102K,DiPorto16}.

However, the count probability distribution function is still missing a model that fits well the observed distribution of galaxies in the three dimensional space. This is the goal of the present paper. We plan to measure the galaxy Counts-in-Cells distribution on the New York University - Value Added Galaxy Catalog \citet{2005AJ....129.2562B} and to assess what is the statistical distribution that fits best the observational frequency distribution. This blind fit of the observational galaxy CiC distribution will be accompanied of a model selection analysis, which will help us to determine the best fitting distribution.

The paper is organized as follows. In Section 2, we describe the data from SDSS \citet{2005AJ....129.2562B} used for the analysis. In section 3, we explain the procedure used to measure the observed Counts-in-Cells (CiC) distribution $f_N(V)$, and describe how we have estimated the errors. In section 4, we introduce different statistical distribution functions to model the observed galaxy CiC distribution. In section 5, we present the observed results and its comparison with the statistical models introduced in Section 4. We summarize our conclusions in Section 6. In the Appendix~\ref{app} we use mock catalogs extracted from the LasDamas simulations \citet{2011AAS...21724907M} to assess the significance of the internal error estimates. 

We use the same values for the cosmological parameters that were used by LasDamas simulation \citep{2011AAS...21724907M} in preparation for the error analysis of Appendix~\ref{app}: $\Omega_{m} = 0.25$, $\Omega_{k} = 0.0$, $\Omega_{\Lambda} = 0.75$ and $H_{0} = 100 h$ km $s^{-1} Mpc^{-1}$.

\section{Data catalogs}\label{cic:data}

The data analyzed in this work is a galaxy sample from the main catalog of the Sloan Digital Sky Survey (SDSS) Data Release 7 (DR7, \cite{2009ApJS..182..543A}). This catalog is provided by The New York University - Value Added Galaxy Catalog \citep{2005AJ....129.2562B}, presented in the next section. In addition to this, we also make use of LasDamas simulation catalog \citep{2011AAS...21724907M} to estimate the uncertainties in $f_V(N)$.

\subsection{The SDSS - New York University - Value Added Galaxy Catalog}\label{nyu}

The NYU-VAGC \citep{2005AJ....129.2562B} is composed by data from the Sloan Digital Sky Survey DR7 \citep{2009ApJS..182..543A} and the 2-Micron All-Sky Survey (2MASS) \citep{1997ASSL..210...25S}, although we only make use of the former one. The SDSS-DR7 mapped one quarter of the entire sky and performed a redshift survey of galaxies, quasars and stars. It consists of a series of three interlocking imaging and spectroscopic surveys, carried out over an eight-year period with a dedicated 2.5m telescope located at Apache Point Observatory in Southern New Mexico.

The NYU-VAGC survey provides us the position and redshift of more than 550.000 galaxies with corrected extinction and K-corrected absolute magnitudes for 8 bands, of which the $u$, $g$, $r$, $i$, and $z$ bands come from the SDSS. In addition, NYU-VAGC catalog also contains a survey geometry catalog, which define the window of the galaxy population and can be operated using the software MANGLE. This survey includes carefully constructed large-scale structure samples useful for calculating power spectra, correlation functions, etc. 

In this work we have used two different samples selected in the $r$ band, corresponding to the DR72 catalogs of the SDSS \citep{2009ApJS..182..543A} included in NYU-VAGC with $r$ band apparent magnitude limit of $17.6$. These two samples are located in the same angular region. 
In this work we reproduce and compare our analysis with data from the LasDamas simulations \citep{2011AAS...21724907M}, and for this reason we select comparable datasets. Our first population is at low redshift, between $0.05$ and $0.106$. The low redshift limit ensures that the sample is within the Hubble flow, and excludes the Coma and Virgo clusters. We take as well a second sample with redshifts between $0.075$ and $0.165$. To determine a suitable absolute magnitude cut, we define a faint limit $M_r$ where the limiting magnitude has been reached to ensure luminosity completeness. in this way, we can assure a similar comoving number density of galaxies within the redshift limits. These limits are $M_r < -20$ for the first population and $M_r < -21$ for the second population. The samples are summarized in Table~\ref{tsdss}. 

\begin{table}
\caption{SDSS selected samples}
\label{tsdss}
\begin{center}
\begin{tabular}{ccccc}
\hline
Sample & Redshift & Magnitude & Density $\bar{n}$ & Galaxies \\
& & $M-5\log(h)$ & $h^{3}$ Mpc$^{-3}$ & \\
\hline
\hline
Pop1 & $0.05$ -- $0.106$ & $M_r < -20$ & $5.7\times 10^{-3}$ & $109046$ \\
Pop2 & $0.075$ -- $0.165$ & $M_r < -21$ & $1.04\times 10^{-3}$ & $84445$ \\
\hline
\end{tabular}
\end{center}
\end{table}

\subsection{LasDamas simulations}\label{data:damas}

In order to obtain reliable estimations of the CiC distribution error bars, we will make use of the multiple realizations of the Large Suite of Dark Matter Simulations (LasDamas) \cite{2011AAS...21724907M,lasdamasweb}, a project that ran a large suite of cosmological N-body simulations that follow the evolution of dark matter in the universe. Results provide us an adequate resolution in many large boxes, rather than a single realization at high resolution. The enormous volume of generated data is appropriate for statistical studies of the distribution of galaxies and halos. The LasDamas simulations are designed to model the clustering of Sloan Digital Sky Survey (SDSS) galaxies in a wide luminosity range, with the goal of assisting in the modeling of galaxy clustering measurements. Specifically, the simulations are used to construct detailed mock galaxy catalogs by placing artificial galaxies inside dark matter halos using the Halo Occupation Distribution (HOD) with parameters fitted to reproduce the galaxy number density and projected correlation function of the respective SDSS galaxy samples. The HOD describes the distribution of galaxies within the dark matter halos. It uses three related properties of the halo model: the probability distribution relating the mass of a dark matter halo to the number of galaxies that form within that halo, the distribution in space of galactic matter within a dark matter halo and the distribution of velocities of galaxies relative to the dark matter within the halo. LasDamas is also designed to reproduce the SDSS DR7 geometry. Altogether, we can see that LasDamas simulations are specially adequate for NYU-VAGC comparison.

Data used consist on two different sets of mocks with different properties, matched to our SDSS samples. One can see them summarized on Table~\ref{lasdamas}. Simulation \textit{Esmeralda} contains 120 realizations and is started at an initial redshift of $z = 99$; while \textit{Carmen} contains 164 realizations and is started at $z = 49$. Mocks are generated with the same initial power spectrum, but a different random seed. 

\begin{table}
\caption{LasDamas selected mock catalogs}
\label{lasdamas}
\begin{center}
\begin{tabular}{ccccc}
\hline
Parent & Redshift & Magnitude & Density & Average nº\\
Simulation & & $M-5\log(h)$ & $h^{3}$ Mpc$^{-3}$ & of particles \\
\hline
\hline
Esmeralda & $0.05$ -- $0.106$ & $M_r < -20$ & $6.01 \times 10^{-3}$ & $121838.9$ \\
Carmen & $0.075$ -- $0.165$ & $M_r < -21$ & $1.11 \times 10^{-3}$ & $81733.8$ \\
\hline
\end{tabular}
\end{center}
\begin{center}
\end{center}
\end{table}

As said, we use LasDamas simulations to estimate the error bars in the CiC distribution. A brief analysis on the reliability of this method and other alternatives can be found in appendix~\ref{app}.

\section{Estimation of the CiC distribution}\label{cic:est}

For our Counts-in-Cells process we use spherical cells in 3-dimensional redshift space with constant radius. These cells are described by their center locations (3 coordinates) and a radius $r$, and randomly distributed in space allowing them to intersect. The higher the number of cells used in the calculation, the most precise will be our calculation of the Counts-in-Cells process. 

\begin{figure*}
\begin{center}
\includegraphics[scale=0.5]{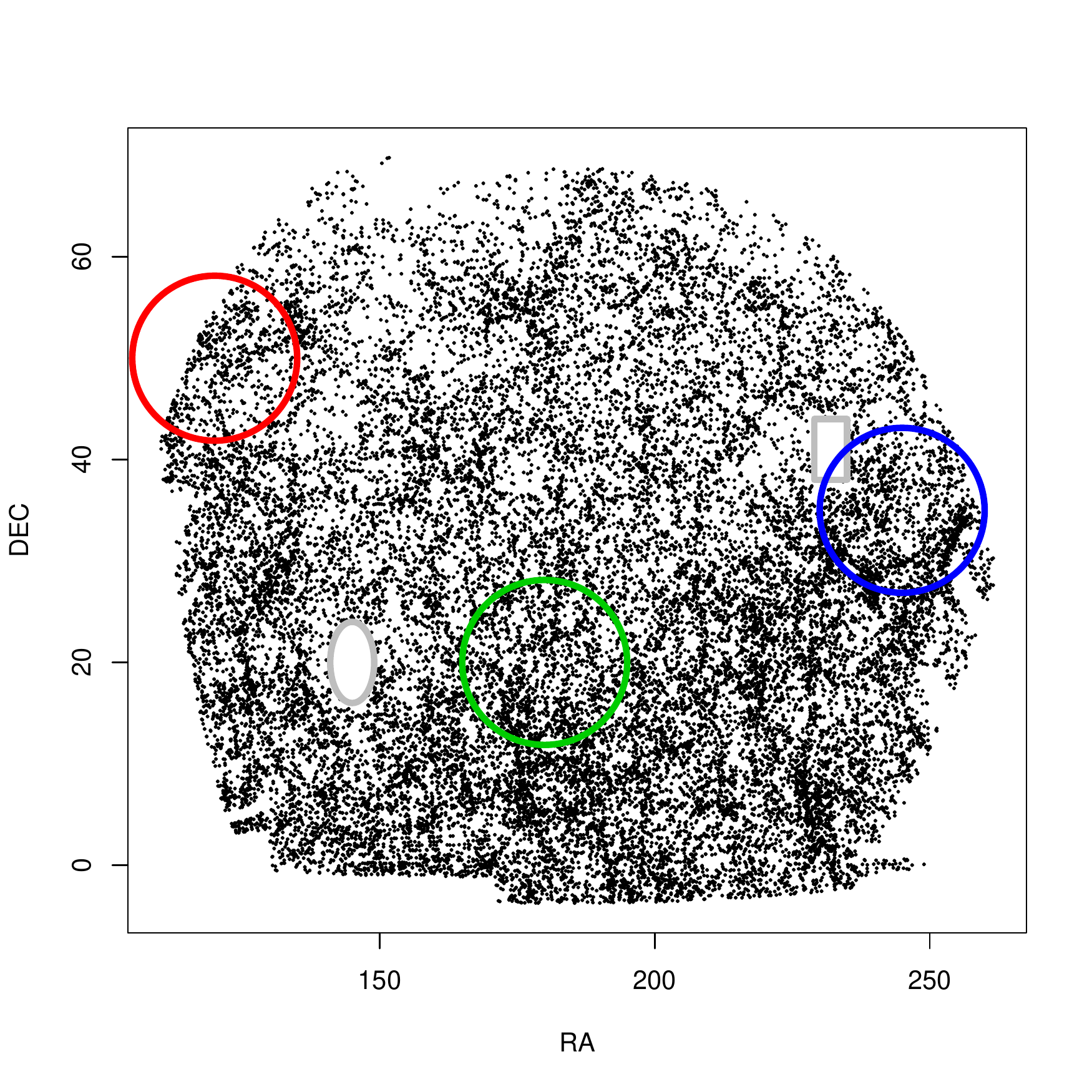}
\includegraphics[scale=0.5]{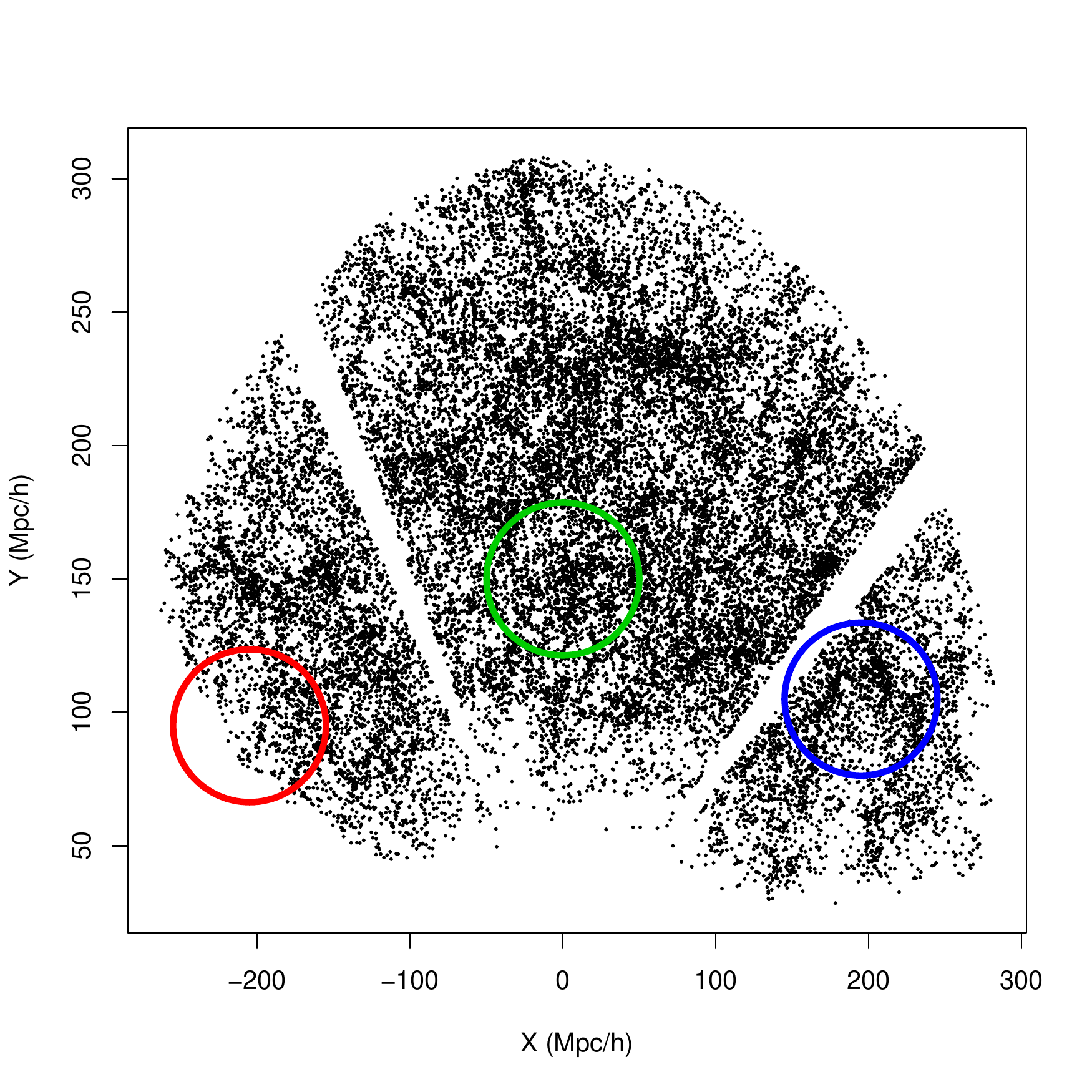}\\
\end{center}
\caption{\label{footrprint} Illustration of the calculation of the counts-in-cells. On the left panel, we show the footprint of about 10\% of the galaxies in the SDSS survey used here (population 1) in right ascension and declination. We can appreciate the irregular sky coverage in the borders outlined by the points. In addition, two masked regions have been added to illustrate the real mask. These masked regions are also illustrated on the right panel where we represent an edge-diagram in Cartesian comoving coordinates. In both panels, the green circle represent a cell which completely falls within the surveyed region, the blue circle represent a cell, still accepted for the counts-in-cells algorithm, having part of its volume outside of the window, but at least 95\% of it lies within the surveyed regions. In this case, counts are compensated to account for the galaxies unseen by the masked regions. Finally, the red circle represents a rejected cell because more than 5\% of its volume falls out of the surveyed region.}
\end{figure*}

A randomly distributed population of cells is generated in the studied survey volume. However, a galaxy survey window can be highly irregular, masked by stars and other objects. This affects not only the distribution of the cells but their effective volume, since even if the center of the cell is inside the window, a significant part of the sphere might be outside. Once the centers of the galaxies are determined, we estimate the effective volume of the cell contained in the window.

For this purpose was created the software MANGLE \citep{2005AJ....129.2562B,2004MNRAS.349..115H,2008MNRAS.387.1391S}. MANGLE performs several operations with complex angular windows, mapping them with sky-projected polygons and allowing one to find the polygons containing a given point on the sphere.

The effective volume of our cells is estimated by Monte Carlo integration, populating a large number of random distributed points in our window and counting the number of points that lie inside our cells. This is exactly a CiC calculation over a random process. Given a window of volume $\nu(W) = V$ and a cell radius $r$, we determine the total number of necessary random points to estimate the effective volume with typical rms uncertainty $a$ as

\begin{equation}
N = 1/a^2 \cdot t ,
\end{equation}

where $t = 4\pi r^3/3V$ is the ratio between the cell volume and the total volume. The number of points in a non masked cell has expectancy $N\cdot t$, and therefore the relative effective volume of each cell is estimated as

\begin{equation}
v_{\rm eff} = N_{\rm obs}/N \cdot t .
\end{equation}

We estimate this effective volume for all cells using a number of random points corresponding to $a = 1\%$. Only cells with $v_{\rm eff} \geq 95\%$ will be accepted.

With our accepted cells we can proceed to perform the Counts-in-Cells. The counting is done using the Euclidian metric and we are interested in knowing how many galaxies are found inside every cell of radius $r$. Then we multiply the number of counts by the inverse of the effective volume to obtain our Counts-in-Cells distribution. With this correction we estimate the number of galaxies unseen by the masking effects or redshift limits assuming uniform distribution of galaxies for cells dimensions. 

The Counts-in-Cells distribution is binned in a histogram of frequencies, each bin containing the number of cells with $n$ galaxies. If we normalize it by the total number of used cells we obtain the probability density function $f_V(N)$ of finding $N$ galaxies in a cell of volume $V$. In our case, we will compute the CiC distribution $f_V(N)$ for our galaxy samples using five different cell radii: $8$, $12$, $16$, $20$ and $24 h^{-1}$ Mpc.

\subsection{Estimation of the errors: LasDamas populations}\label{jack}

The estimation of the uncertainty in our CiC measurements are done using the presented LasDamas populations. These samples have been selected so they can be compared with the SDSS populations. Using our accepted cells and their volume correction, we repeat the CiC calculations over LasDamas realizations, obtaining a reliable sampling of the distribution of cells containing a given number of galaxies. We will use the standard deviation of these quantities as an error estimation for our CiC distribution over the SDSS populations. We compute the mean and the variance with

\begin{eqnarray} \label{mean}
\bar{x}_{j} &=& \frac{1}{N_s}\sum^{N_s}_{i=1}f_{V}^{j}(i) \nonumber \\
\sigma_{j}^2 &=& \frac{1}{N_s-1}\sum^{N_s}_{i=1}(f_{V}^{j}(i) - \bar{x}_{j}^{j}(i))^2
\end{eqnarray}
where $N_s$ is the number of LasDamas samples and $f_{V}^{j}(i)$ is the $i$th realization distribution evaluated at the number of galaxies per cell $N=i$. We opted for using these quantities instead of errors estimated from intrinsic methods (i.e, the delete-one Jackknife method, \cite{2009MNRAS.396...19N} or the bootstrap method, \cite{efron1994introduction}) due to an overestimation of the found errors. An analysis of this issue can be found in appendix~\ref{app}.

The standard deviation values $\sigma_j$ found with LasDamas realizations provide error estimation for a large quantity of number of galaxies per cell $N$. However, for radii $20$ and $24 h^{-1}$ Mpc we will see how, for every realization, no cells where found with a certain small number of galaxies inside. These values have $f_V(N) = 0$ with no estimation of their uncertainty, for which we assume the closest positive $\sigma_j$ value. This is specially important for the case of the void-probability function, the probability of finding an empty cell ($f_V(0)$). These value might affect the fitting results (section~\ref{sec:fitting}) and must be included in the best fit estimation. For large values of $N$ this problem can be neglected.

As we will see in section~\ref{sec:fitting}, we will make use of these standard deviations as weights for our $\chi^2$ fitting procedure (see eq.~\ref{lsd}). This implies assuming a Gaussian distribution for the observed abundances of cells containing $N$ galaxies. However, the Shapiro-Wilk test for Gaussianity \citep{royston1982extension} reveals how these distributions are not Gaussian when $f_V(N)$ is close to 0. We have not been able to find any known probability density satisfactorily modeling these distributions at all scales and further analysis is needed. As explained in section~\ref{sec:fitting} we perform the $\chi^2$ fit twice, first using all the obtained $\sigma_i$ values and then only with those satisfying Gaussianity. Since most of the non Gaussian distributed errors correspond to high values of $N$, where the CiC distribution describe an asymptotic tail, negligible differences were found between both best fit parameters. However, the number of used bins can be crucial for the $\chi^2$ value and the parameters error bars. The chosen solution consists in increasing the size of those bins where Gaussianity cannot be assured including the cell counts from the first Gaussian bin on the right. This creates a wide bin for the smaller $N$ values in large cell radius samples, and another wide bin for the right tail. Altogether, Gaussianity of all bins can be assured to be above $0.9$ in the Shapiro-Wilk test.

\section{Frequency distribution models}
\label{sec:fdf} 

In this section we present the probability distribution functions used to model the observed Counts-in-Cells distributions from the SDSS. Some of these functions have been commonly used in the cosmological literature to fit the CiC distribution, and some are an original contribution of this work. It is therefore of high interest to discriminate the best fitting model and their best fitting parameter values. Despite the information contained in a CiC distribution does not fully characterize the galaxy distribution, it is a useful constraint for N-body simulations or models of the galaxy distribution. Now we present the used probability density functions. 

\subsection{Gravitational Quasi-Equilibrium Distribution}
\label{gqed}

The GQED \citep{1984ApJ...276...13S} is derived from a thermodynamical description of the galaxy fluid, though it can be also derived from the statistical mechanics  \citep{2002ApJ...571..576A}. One of its principals assumptions is to accept that the galaxy accretion is succeeded through series of quasi-equilibrium states, a basic condition to start the thermodynamical approximation. This is possible if we assume an infinite quantity of galaxies in the universe. With this assumption we say clusters and large structures remain stable along large cosmological times before changing into a new quasi-equilibrium state. 

We will assume as well that galaxies are formed without dynamical interactions with the exterior and then they interact gravitationally as punctual masses with the rest of the universe. As the clusters dimensions are smaller than the curvature radius of the universe and the velocities involved are much lower than the speed of light, we can assume Newtonian gravity with potential $\phi = r^{-1}$.

After these assumptions, the GQED can be derived as the pdf

\begin{equation}
f_V(N) = \frac{\bar{N}(1-d)}{N!}[\bar{N}(1-d) + N\cdot d]^{N-1}e^{-[\bar{N}(1-d) + N \cdot d]}
\end{equation}
where the expected value of $N$ is the product of the cell volume by the mean density of the galaxies, $\bar{N} = \bar{n}V$. 

Details in the deduction and properties of this distribution can be fully consulted in \cite{1984ApJ...276...13S,1996ApJ...470...78S} and \cite{1995MNRAS.274..213S}.

Parameter $d$ (named $b$ in \cite{2011ApJ...729..123Y}) allows us to study both physical and statistical properties of this distribution. For the limit $d=0$ we have a random distribution, where galaxies are uniformly distributed. For big scales, where fluctuations are small, the density function becomes Gaussian. Since this $d$ value can be determined from physical magnitudes of the galaxies, we have a free parameter model of the galaxy distribution. As we find in \cite{2002ApJ...571..576A}, $d$ represents a measure of the state of aggregation and we can express it as 

\begin{equation}
d = \frac{3/2(Gm^2)^3\bar{n}T^{-3}}{1+3/2(Gm^2)^3\bar{n}T^{-3}}
\end{equation}
This quotient relates $d$ to galaxy mass $m$, mean density $\bar{n}$, galaxies kinetic temperature $T$ and gravitational constant $G$.

In the fitting process we will show the resulting parameters obtained directly from the CiC distribution (equation~\ref{impb}) and parameters $\bar{N}$ and $d$ when both are fitted. The calculation of this distribution and the function Gamma for large values of $N$ includes considerable difficulties, since the evaluation of the factorial function requires dealing with large numbers and needs a high precision to avoid the propagation error along the loops. As we will see in the following sections, this is the case for large cell radius distributions. This difficulty has been overcome using continuous fractions as representation of real numbers with infinite (i.e., arbitrarily large) precision and exploiting the model of lazy evaluation in the Haskell programming language \citep{haskell98}, which allows us to perform heavy calculations and it is easy to integrate in C++.

In addition, we will compare these results with the estimated values of $\bar{N}$ and $d$ obtained directly from the CiC distribution. As explained in \cite{2011ApJ...729..123Y}, the dependence of $d$ on $V$ can be obtained empirically from the variance of the number of counts in a cell of volume $V$ ($\langle(\Delta N)^{2}_{V}\rangle$) and the mean $\bar{N}$.

\begin{equation}\label{impb}
d(V) = 1 - (\bar{N}\bar{\xi}_{2}(V) + 1)^{-1/2}
\end{equation}
where $\xi_2$ is the 2-point correlation function.

\subsection{Negative Binomial Distribution}\label{nbd}

The Negative Binomial Distribution (NBD) was firstly proposed in the cosmological context by \cite{1983PhLB..131..116C} and derived later by \cite{1992MNRAS.254..247E}. With this model we study the probability of distributing $N$ galaxies in $m$ disconnected boxes. The probability of finding a galaxy in one of these boxes is proportional to the number of galaxies already located in the box. Expressing the probability function in terms of galaxies per cell instead of its conventional form, we have:

\begin{equation}
f_{V}(N) = \frac{\Gamma(N+\frac{1}{g})}{\Gamma(\frac{1}{g})N!}\frac{\bar{N}^N(\frac{1}{g})^{\frac{1}{g}}}{(\bar{N}+\frac{1}{g})^{N+\frac{1}{g}}}
\end{equation}
where $\Gamma$ is the gamma function.
Similarly as we had with the value $d$ for the GQED, $g$ is also an aggregation parameter. We can obtain it theoretically for the NBD with

\begin{equation}\label{gNBD}
g = \frac{\langle(\Delta N)^2\rangle - \bar{N}}{\bar{N}^2},
\end{equation} 
In this model, $g$ depends both on the volume and on the shape of the cells.

This function is widely used in statistics and was firstly introduced in the cosmological context as a phenomenological model without physical explanation. Some authors \citep{betancort2000generalized} claim that this function appear as a limit case in a variety of processes, in particular \cite{1983PhLB..131..116C} found that approximates rather well the distribution of Zwicky clusters. Others, however, found that violates the second law of thermodynamics \citep{2011ApJ...729..123Y,1996ApJ...460...16S}. However, it provides a fair agreement with the observational distribution and it is thought to be related with the hierarchical universe properties. A deeper study of the implications of using such distribution suggests that the NBD assumes galaxies to be formed where there is already a galaxy cluster. Hence, this model does not take infall into account, but can describe the case where galaxies form from the merging of less massive objects. As before, we will show the resulting parameters obtained from equation~\ref{gNBD} and parameters $\bar{N}$ and $g$ when both are fitted. Similarly than the GQED distribution, the NBD involves the calculation of factorials for large N values and infinite precision, which requires versions of these functions parameterized with the precision required, as we have already noticed. We have also compared the results of Haskell programming mentioned above 
with the use of the Stirling approximation for the factorials of big numbers, by means of the MAXIMA program \footnote{http://maxima.sourceforge.net/}. The differences of the results are negligible.

\subsection{Log Normal Distribution}\label{lnd}

The Log Normal Distribution (LND) was firstly used by \cite{1934ApJ....79....8H} but it was not formally proposed until \cite{1991MNRAS.248....1C}. It was one of the first fully described stochastic models applied to the distribution of matter density. This model allows us to calculate many complex properties of the CiC distribution. 

While the Gaussian distribution provides us a valid description of linear and weak density perturbation fields, the Log Normal distribution represents the same case for the non-linear regime and is one of the few non-Gaussian random fields for which interesting properties are calculable analytically. In \cite{1991MNRAS.248....1C} one may find further motivation for this model, such as agreements with observational data. This has turned the Log Normal distribution into one of the most well known and widely applied models \citep{2005MNRAS.356..247W,2010MNRAS.403..589K,2001ApJ...561...22K}.

Given a Gaussian random field $X(r)$, the probability of finding a certain number density from our field in a given position is given by $X \sim N(\mu,\sigma^2)$. The Log Normal field is obtained by applying a non-linear transformation of the Gaussian field: $Y(r) = \exp{(X(r))}$, where the new 1-point probability function in the field $Y$ is

\begin{equation}\label{lneq}
f_{1}(y) = \frac{1}{\sigma\sqrt{2\pi}}\exp{\left(-\frac{(\log y - \mu)^2}{2\sigma^2}\right)}\frac{1}{y}
\end{equation}

This is the Log Normal variable $Y \sim \Lambda(\mu,\sigma^2)$ pdf, with the same parameters as the original Gaussian variable. In this work we make use of an expression adapted to the cosmological scenario \citep{2016JCAP...03..005A}

\begin{equation}\label{lnpdf}
f(\Delta) = \frac{1}{\sqrt{2\pi H_0}}\exp{\left(-\frac{y^2}{2H_0}\right)}\frac{1}{\Delta}
\end{equation}
where

\begin{equation}
\begin{split}
\Delta &= N/\bar{N}\\
H_0 &= \log{(1+C)}\\
y &= \log{\left(\Delta\sqrt{1+C}\right)}
\end{split}
\end{equation}
and $C$ is the variance in the matter distribution, which we expect to be roughly $C = \sigma_8^2$ for cell radius $r = 8 h^{-1}$ Mpc at $z=0$. This distribution has proved efficient in the description of the dark matter density field \citep{2016JCAP...03..005A,2017MNRAS.466.1444C}, understood as a local non-linear transformation of the initial energy density distribution. However, the galaxy distribution studied in this work presents a bias with respect to the dark matter distribution, and the Log Normal distribution must be modified conveniently. This bias parameter $b$ can be introduced in the probability density function as shown in \cite{1999ApJ...520...24D}.

\begin{equation}\label{lnbpdf}
f(\Delta) = \frac{1}{\sqrt{2\pi H_0}}\frac{\exp{(-\frac{1}{2}\frac{y^2}{H_0})}}{\Delta+b-1}
\end{equation}
where the above expressions are conveniently modified into

\begin{equation}
\begin{split}
H_0 &= \log{(1+C_b)}\\
y &= \log{\Big((\Delta+b-1)\frac{\sqrt{1+C_b}}{b}\Big)}
\end{split}
\end{equation}
and $\Delta$ remains the same. With this new distribution we expect improved fittings of the galaxy CiC distribution. 
In this case, we assume the simplest case of a scale-independent and linear bias, so $\delta_g = b \delta$, where $\delta_g$ and $\delta$ are the contrasts of the galaxy number density and the matter density, respectively.

\subsubsection{Shot noise correction}
\label{sssec:shotnoise}

As explained above, Log Normal distribution is valid to describe a continuous field, in the cosmological context this is either the dark matter or the galaxy density fields. In our case, therefore, we need to introduce a discrete sampling of that underlying continuous field in order to compare to the observed counts-in-cells. This discrete sampling introduces a modification of the distribution known as shot noise. It is more important at smaller scales, where the typical number of galaxies per cell is low.

We correct for this shot noise term assuming the local Poisson process model \citep{layzer1956new, 1980lssu.book.....P}. This has been done in the literature both at the level of the moments of the distribution \citep{Gaztanaga02,Angulo08,BelMar14,Bel14} or, as we are dealing in this paper, at the level of the probability distribution function \citep{DiPorto16,2016A&A...588A..51B,Hoffmann17}. Following these authors the modified probability of galaxy counts, $f(N)$, can be expressed as.

\begin{equation}\label{newln}
f(N) = \int^{+\infty}_{-1} f(\Delta)P(N|\Delta)d\Delta
\end{equation}

here $P(N|\Delta)$ is  the conditional probability of finding $N$ galaxies in a Poisson process with density $\Delta$. Its expression is

\begin{equation}
P(N|\Delta) = \frac{(\bar{N}\Delta)^{N}e^{-\bar{N}\Delta}}{\bar{N}!}
\end{equation}

Althought in this work we will limit ourself to the use of a Poisson distribution, different alternatives can be used to correct from shot noise effects. In the following sections, we will use equation~\ref{newln} as our Log Normal distribution $f_V(N)$ (with or without bias correction).

\section{Results}\label{res}

Counts-in-Cells calculations have been performed making use of at least as many cells as galaxies in the sample, which we consider a sufficient condition \citep{2011ApJ...729..123Y}. In addition, nearly 50\% of cells are rejected by mask effects, so we have to double the number of tested cells. In Table~\ref{fitcic1} one can find the numbers of used cells for each SDSS sample and LasDamas realizations under column Cells. 
The Counts-in-Cells observed probability density functions $f_V(N)$ are shown in Figs.~\ref{fig1} and~\ref{fig2} together with the best fits of the different distributions. The curves show the expected probabilities for our chosen samples and radii \citep{2011ApJ...729..123Y}, with higher values of $\bar{N}$ 
for bigger cells. Used bins variate their width depending on the CiC population. We have used the Freedman-Diaconis rule \citep{freedman1981histogram}, which fixes the width $m$ of the bin depending on the number of observations and the spread of the population:

\begin{equation}
m = 2 \frac{IQR(X)}{n^{1/3}}
\end{equation}
where $IQR(X)$ is the interquartile of population $X$ and $n$ is the number of objects in $X$. The resulting quantity has been rounded, resulting on $m=1$ for cell radii $8$ and $12 h^{-1}$ Mpc, $m=3$ for radius $16 h^{-1}$ Mpc and $m=6$ for radius $24 h^{-1}$ Mpc in Population 1. In Population 2, all bins have width $m=1$, except for radius $24 h^{-1}$ Mpc, where $m=2$.

\subsection{Fitting the results to a distribution function}\label{sec:fitting} 

In this section we proceed to fit the probability distribution functions defined in section~\ref{sec:fdf} to our Counts-in-Cells observed distributions. We calculate the $\chi^2$ goodness of fit 

\begin{equation}\label{lsd}
\chi^2 = \sum_{N=0}^{N_{max}}\frac{(f_{V}^{obs}(N)-f_{V}(N,\theta))^2}{\sigma_N^2}
\end{equation}
between the observed distribution and the theoretical distribution as a qualitative measure, where $N_{max}$ is the largest number of galaxies in a cell. The fitting will make use of the diagonal values $\sigma^2$, ignoring correlations between bins. This estimation might underestimate the obtained best fit $\chi^2$ values and the parameters $1\sigma$ error bars, however, we have checked that this decision does not affect the best fit values. After generating samples of Poisson-distributed points in the used masks with the densities of Table~\ref{tsdss}, we performed the described effective volume estimation method. The analytical Poisson distribution for the used densities are within the uncertainties of the obtained results and thus, we are confident of our methods. Vector $\theta$ contains the parameters of the theoretical distribution.

\begin{figure*}
\begin{center}
\includegraphics[scale=1.0]{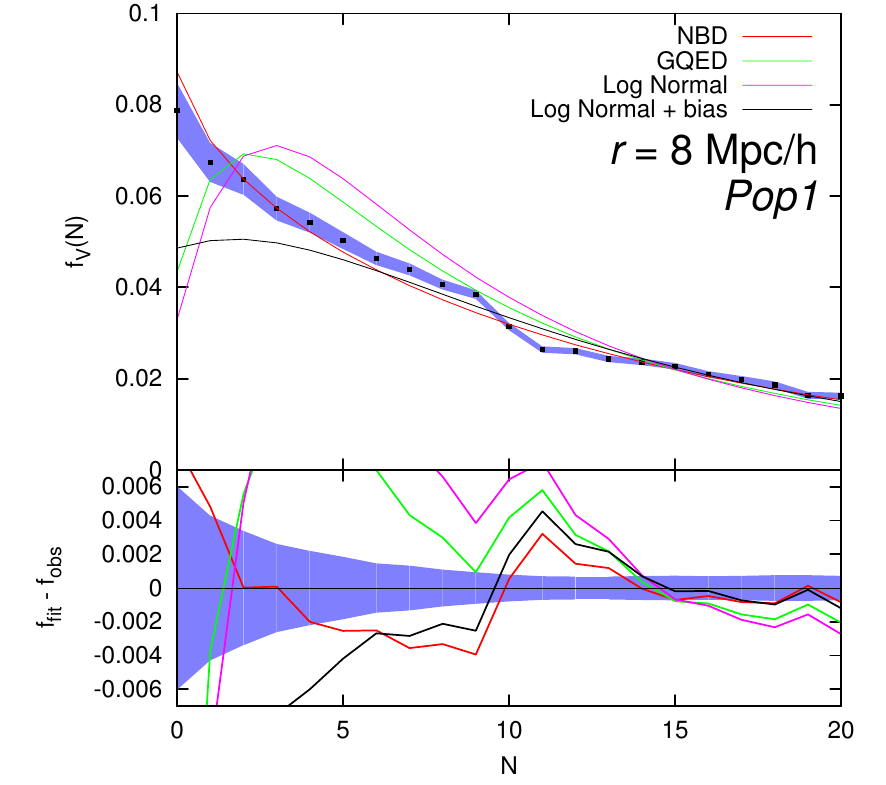}
\includegraphics[scale=1.0]{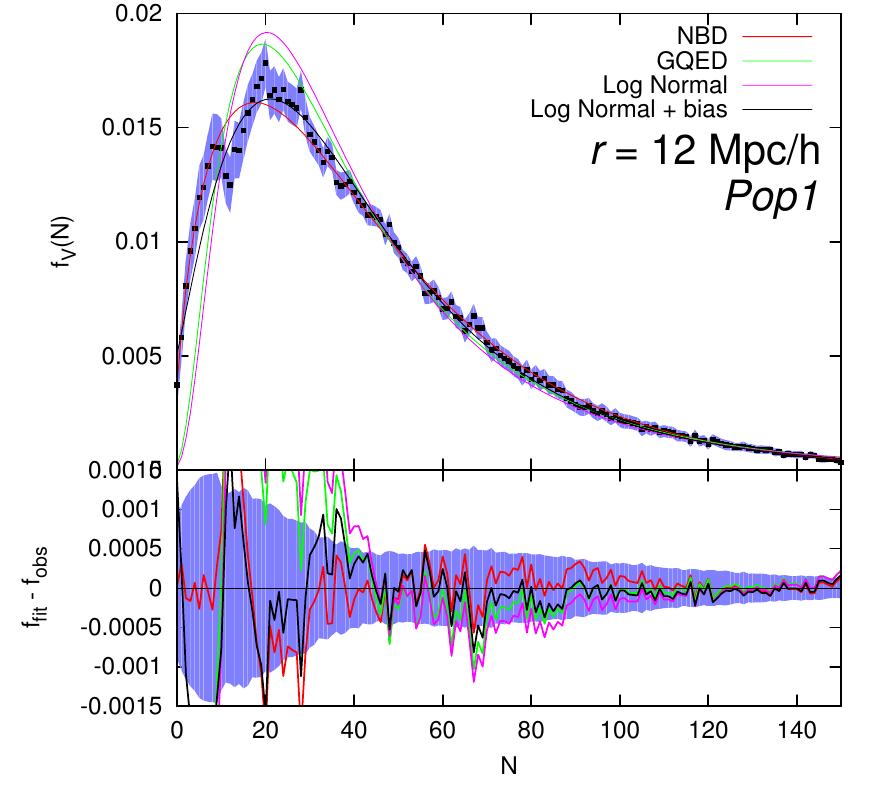}\\
\includegraphics[scale=1.0]{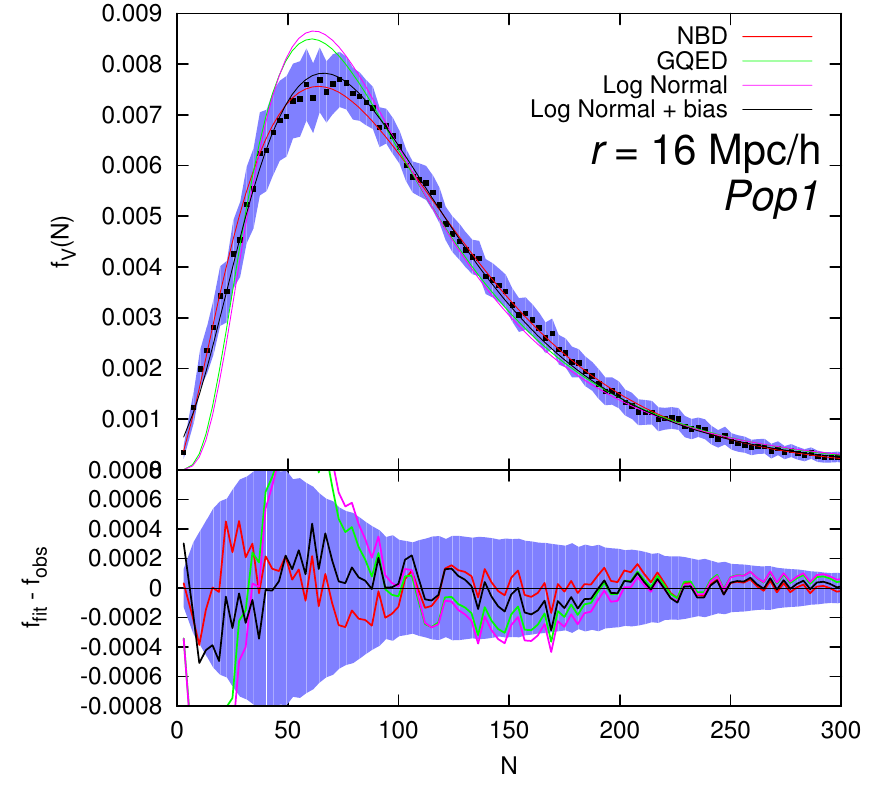}
\includegraphics[scale=1.0]{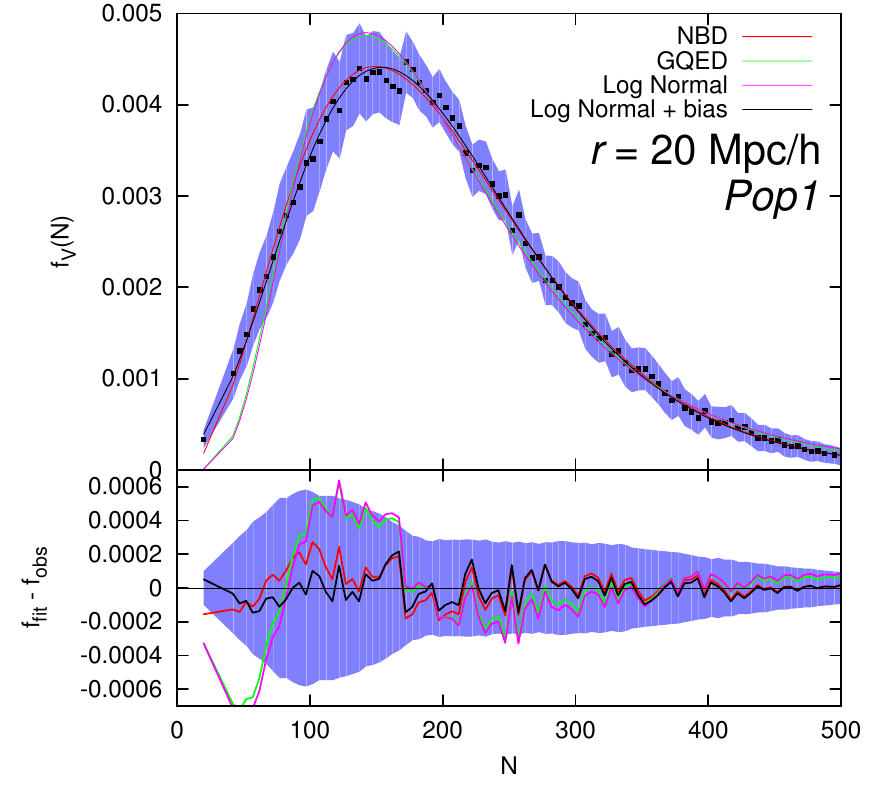}\\
\includegraphics[scale=1.0]{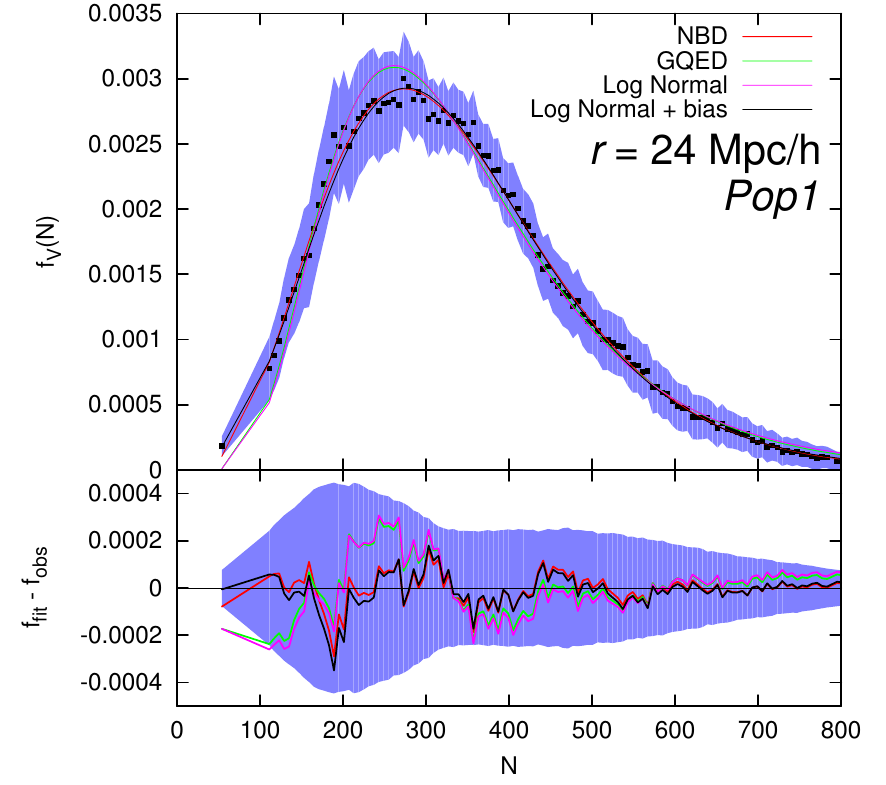}
\end{center}
\caption{\label{fig1}Count-in-cells results for population 1. $f_V(N)$ CiC distribution function with best fit models and error bars. Top to bottom and left to right: radii $8$, $12$, $16$, $20$ and $24 h^{-1}$ Mpc. Best fit plots for bins with Gaussian errors. Boxes: NBD residuals, GQED residuals, Log Normal distribution residuals, Log Normal with bias distribution residuals.}
\end{figure*}

\begin{figure*}
\begin{center}
	\includegraphics[scale=1.0]{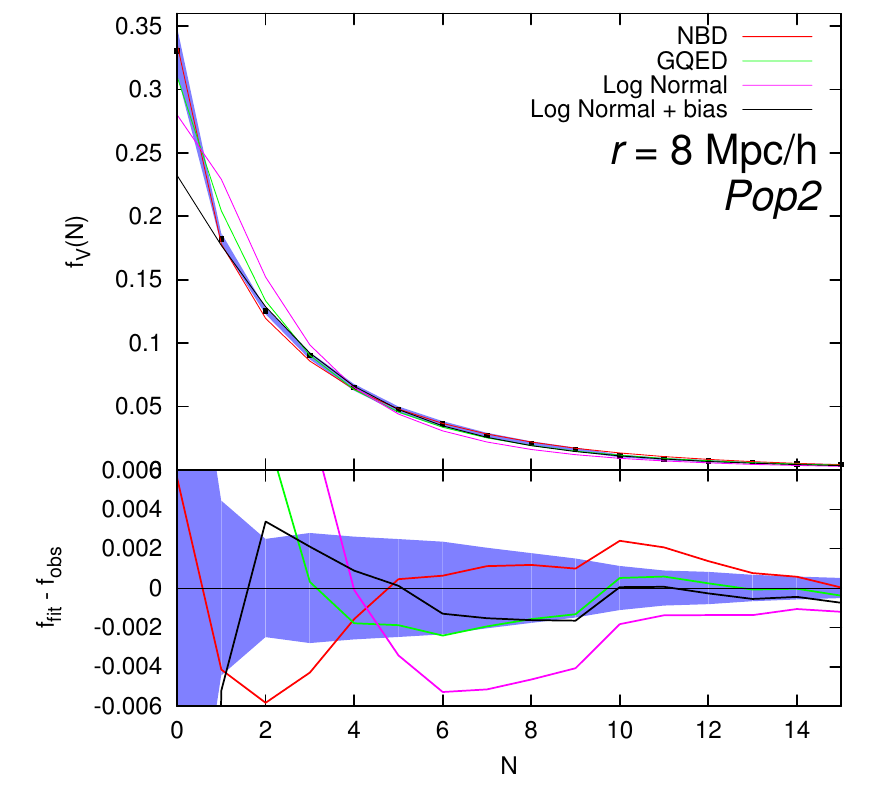}
	\includegraphics[scale=1.0]{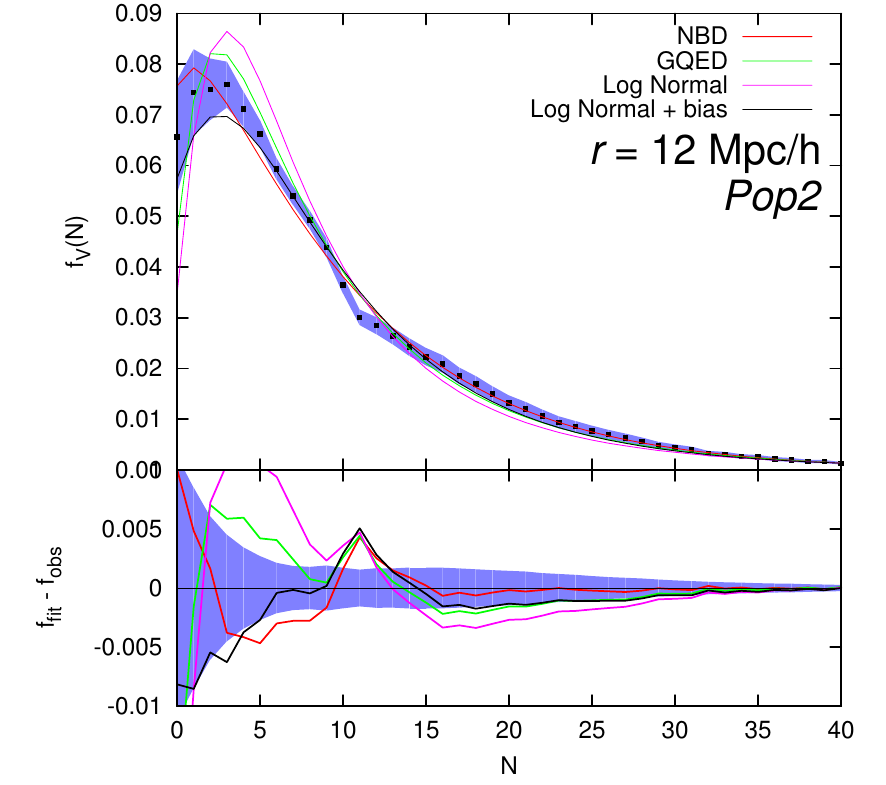}\\
	\includegraphics[scale=1.0]{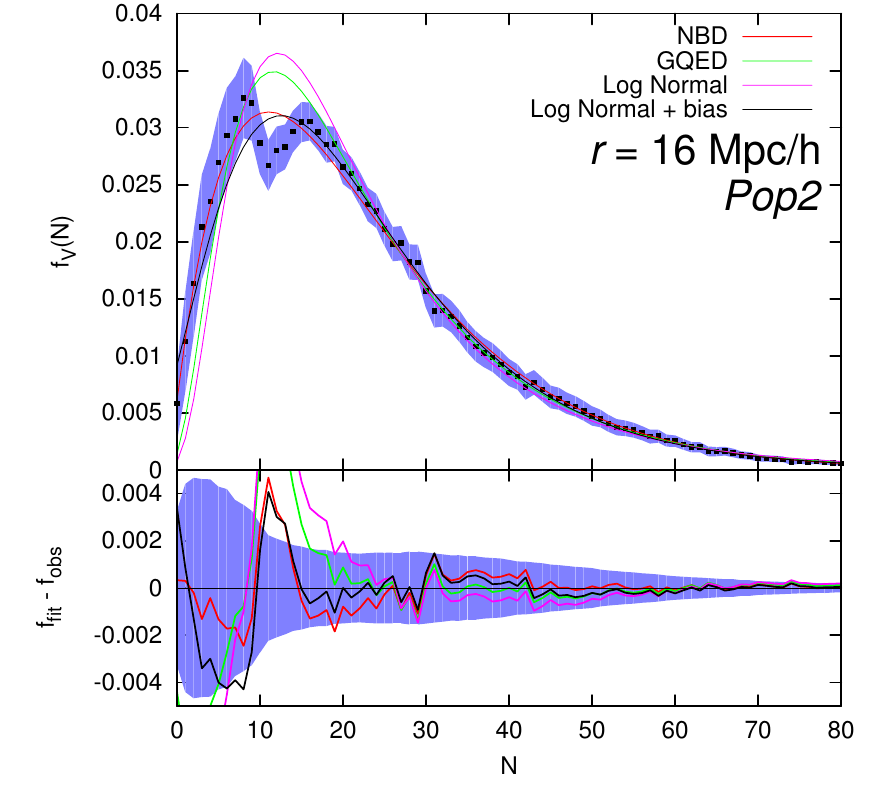}
	\includegraphics[scale=1.0]{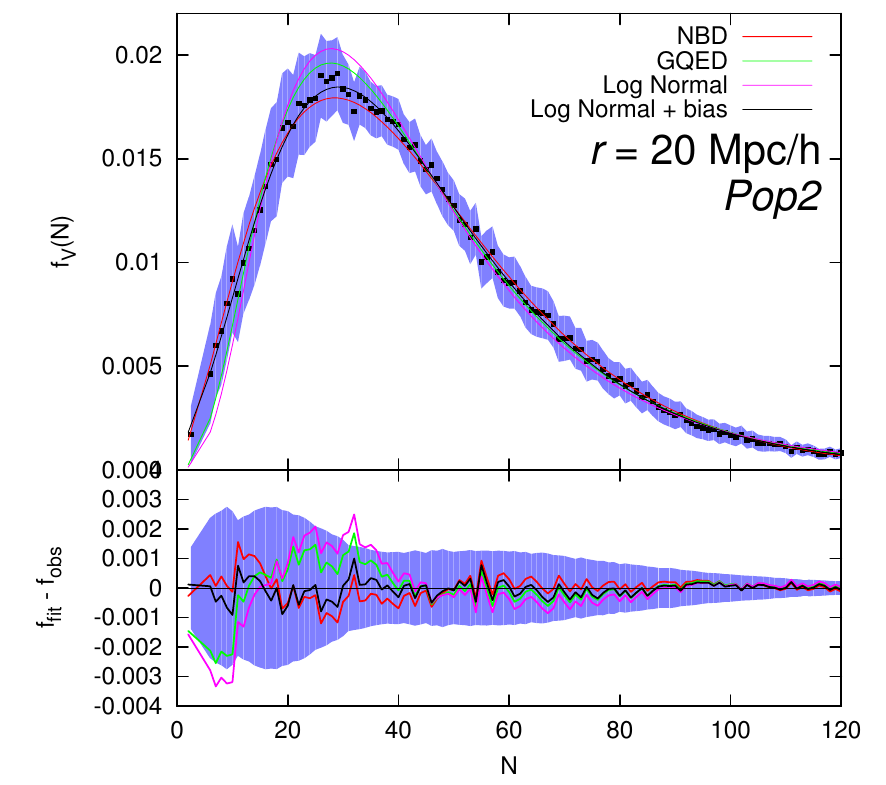}\\    
	\includegraphics[scale=1.0]{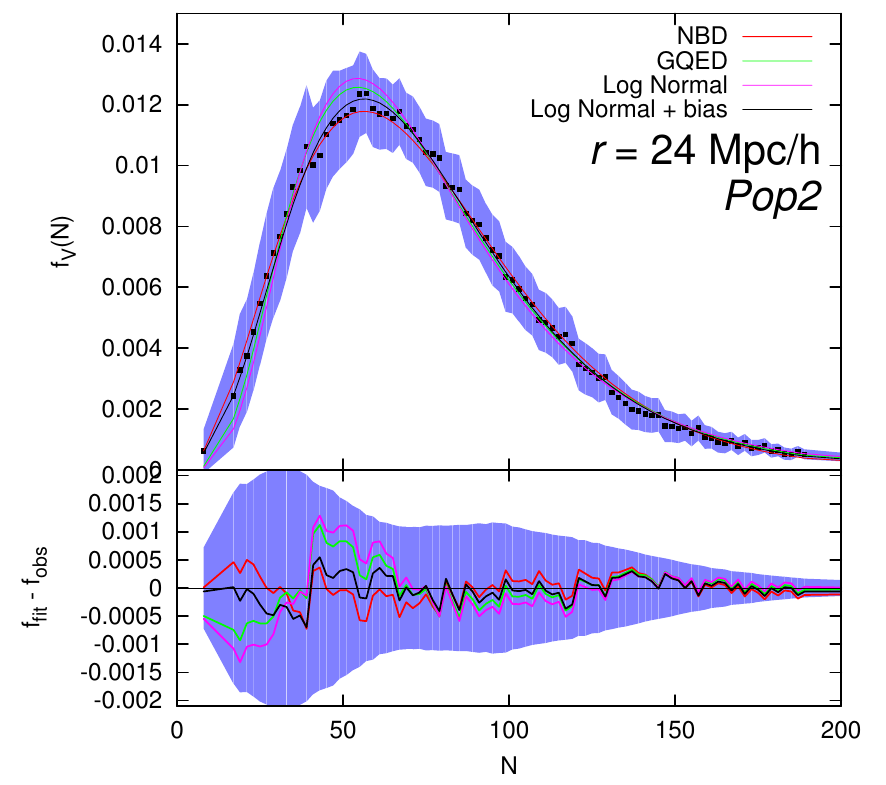}
\end{center}
\caption{\label{fig2}Count-in-cells results for population 2. $f_V(N)$ CiC distribution function with best fit models and error bars. Top to bottom and left to right: radii $8$, $12$, $16$, $20$ and $24 h^{-1}$ Mpc. Best fit plots for bins with Gaussian errors. Boxes: NBD residuals, GQED residuals, Log Normal distribution residuals, Log Normal with bias distribution residuals.}
\end{figure*}

As introduced in section~\ref{jack}, we have performed our fitting analysis over an observational distribution function with adaptive bin width, with wider bins for those values where $f_V(N)$ was close to zero and the Gaussianity of its error distribution could not be assured. The solution consisted on merging these bins with the closest Gaussian bin on the right. The best fit solution for the full curve analysis for each population can be found respectively in Figs.~\ref{fig1} and~\ref{fig2}.

We show in Table~\ref{fitcic1} the values of $\theta$ and $\chi^2$ obtained by performing a blind fit where all parameters are allowed to variate. Fittings have been performed using our own developed routines and also by means of the non-linear least-squares fitting MPFIT routines in IDL \cite{mpfit}. Differences between both procedures are negligible. The numbers reported here are the ones by our own routines.

\begin{table*}[t]
\caption{Counts-in-Cells best fit $f_{V}(N)$ - All parameters free}
\begin{center}
\begin{tabular}{lrc|cccc|ccccc}
\hline
\multicolumn{3}{c}{Sample} & \multicolumn{4}{c}{NBD} & \multicolumn{5}{c}{GQED}\\
\hline
Pop & Cells & $r$ & $\bar{N}$ & $g$ & $\chi^2$ & $dof$ & $\bar{N}$ & $d$ & $\chi^2$ & $dof$\\ 
\hline
\hline
1 & 195079 & 8		 & 	$13.08 \pm 0.09$ & $1.131 \pm 0.014$ & $124.56$ & $124$ & $13.12 \pm 0.08$ & $0.761 \pm 0.0008$ & $417.87$ & $124$\\
	 & 167367 & 12		 & 	$44.1 \pm 0.2$ & $0.59 \pm 0.006$ & $97.75$ & $215$ & $45.67 \pm 0.18$ & $0.8177 \pm 0.0008$ & $326.54$ & $215$\\
	 & 144938 & 16		 & 	$102.5 \pm 0.8$ & $0.373 \pm 0.005$ & $17.04$ & $120$ & $105.6 \pm_{0.8}^{0.6}$ & $0.8482 \pm_{0.0017}^{0.0008}$ & $115.66$ & $120$\\
     & 123972 & 20		 & 	$200.1 \pm 1.6$ & $0.252 \pm_{0.005}^{0.007}$ & $11.01$ & $100$ & $206.4 \pm 1.7$ & $0.87 \pm 0.0017$ & $60.2$ & $100$\\
     & 215408 & 24		 &  $338 \pm 3$ & $0.191 \pm 0.005$ & $6.71$ & $115$ & $347 \pm 3$ & $0.8812 \pm 0.0018$ & $24.65$ & $115$\\
\hline
2 & 218339 & 8		 &  $2.81 \pm 0.02$ & $1.53 \pm 0.03$ & $35.79$ & $34$ & $2.75 \pm 0.04$ & $0.575 \pm 0.003$ & $51.69$ & $34$\\
	 & 196647 & 12		 &  $9.34 \pm 0.07$ & $0.848 \pm 0.017$ & $34.04$ & $60$ & $9.36 \pm 0.09$ & $0.672 \pm 0.003$ & $59.87$ & $60$\\
     & 177485 & 16		 &  $22.54 \pm 0.18$ & $0.488 \pm 0.01$ & $29.17$ & $96$ & $23.02 \pm 0.18$ & $0.715 \pm 0.003$ & $78.22$ & $96$\\
     & 161357 & 20		 &  $43.7 \pm 0.3$ & $0.336 \pm 0.007$ & $18.85$ & $132$ & $44.6 \pm 0.4$ & $0.753 \pm_{0.003}^{0.002}$ & $22.9$ & $132$\\
     & 147554 & 24		 & 	$75 \pm 0.6$ & $0.241 \pm 0.008$ & $9.62$ & $87$ & $76.6 \pm 0.8$ & $0.777 \pm 0.004$ & $8.81$ & $87$\\
\hline
\multicolumn{3}{c}{} & \multicolumn{4}{c}{Log Normal} & \multicolumn{5}{c}{Log Normal + bias}\\
\hline
 &  &  & $N$ & $C$  & $\chi^2$ & $dof$ & $\bar{N}$ & $C_b$ & $b$ & $\chi^2$ & $dof$\\
\hline
\hline
1 & 195079 & 8 & $13.17 \pm 0.09$ & $1.66 \pm 0.02$ & $1097.83$ & $124$ & $12.53 \pm 0.1$ & $0.71 \pm 0.03$ & $1.42 \pm 0.02$ & $238.17$ & $123$\\
	 & 167367 & 12		 & 	$46.6 \pm 0.2$ & $0.743 \pm 0.011$ & $542.29$ & $215$ & $44.5 \pm 0.3$ & $0.4005 \pm 0.0005$ & $1.297 \pm 0.009$ & $96.42$ & $214$\\
	 & 144938 & 16		 & 	$107.4 \pm 0.9$ & $0.447 \pm 0.011$ & $157.18$ & $120$ & $103.9 \pm 0.7$ & $0.241 \pm 0.017$ & $1.29 \pm 0.03$ & $21.2$ & $119$\\
    & 123972 & 20		 & 	$213 \pm 2$ & $0.308 \pm 0.011$ & $71.88$ & $100$ & $200 \pm 2$ & $0.113 \pm 0.018$ & $1.52 \pm 0.1$ & $7.08$ & $99$\\
     & 215408 & 24		 &  $351 \pm 3$ & $0.215 \pm 0.008$ & $30.03$ & $115$ & $338 \pm 4$ & $0.09 \pm 0.015$ & $1.48 \pm 0.15$ & $6.6$ & $114$\\
\hline
2 & 218339 & 8		 &  $2.16 \pm 0.05$ & $5.6 \pm 0.3$ & $394.31$ & $34$ & $2.69 \pm 0.04$ & $1.1 \pm 0.07$ & $1.31 \pm 0.03$ & $50.03$ & $33$\\
	 & 196647 & 12	&	 $9.65 \pm 0.2$ & $1.26 \pm 0.05$ & $201.82$ & $60$ & $9.18 \pm 0.13$ & $0.54 \pm 0.04$ & $1.35 \pm 0.05$ & $51.8$ & $59$\\
    & 177485 & 16		 & $23.5 \pm 0.2$ & $0.547 \pm 0.014$ & $156.64$ & $96$ & $22.4 \pm 0.3$ & $0.252 \pm 0.03$ & $1.45 \pm 0.07$ & $29.54$ & $95$\\
     & 161357 & 20		 &  $45.3 \pm 0.4$ & $0.381 \pm 0.011$ & $45.87$ & $132$ & $44.1 \pm 0.4$ & $0.21 \pm 0.02$ & $1.3 \pm 0.07$ & $10.54$ & $131$\\
     & 147554 & 24		 &  $77.6 \pm 0.9$ & $0.269 \pm 0.012$ & $13.34$ & $87$ & $75.9 \pm 1$ & $0.16 \pm 0.04$ & $1.25 \pm 0.1$ & $6.03$ & $86$\\
\hline
\end{tabular}
\end{center}
\label{fitcic1}
{\bf Notes.}
Best fit parameters from the Counts-in-Cells distribution for our four probability functions. Different bin widths have been used, with wider bins for larger cells radii, resulting on different degrees of freedom for the $\chi^2$ fitting. See section~\ref{res} for detailed information. Column \textit{Cells} corresponds to the used number of cells in each case. 
\end{table*}

We use the $\chi^2$ values to discriminate the best fitting distribution. These values can be compared with results in Figs.~\ref{fig1} and~\ref{fig2}. Top box shows the observational and best fit distributions. In the bottom boxes we can see the residuals, the difference between the best-fit distributions and the observed one (X axis). As expected, curves outside the error bars have higher values for $\chi^2$.

Regarding the best fit distribution, the Negative Binomial Distribution generally performs the best fit results with lower values of $\chi^2$. This is specially clear for radius $8 h^{-1}$ Mpc. However, we must notice that, for higher radii, the Log Normal distribution with bias obtains remarkably good values as well, specially the higher radii of Population 2. 
At small scales, the Log Normal distribution with bias show a worse fit than the NBD. 
This may be an indication that the correction introduced in Sect.~\ref{sssec:shotnoise} does not completely correct for the shot noise term, or that this underlying distribution fails to describe the density field at these scales.

The Gravitational Quasi-Equilibrium Distribution generally performs poorer fittings than the NBD or the LN with bias distribution. But its $\chi^2$ values are often close to those obtained by the LN. Finally, the LN distribution, without the bias modification, performs, as expected, worse fittings than any other distribution, clearly showing its unsuitability to describe the galaxy distribution, although it is rather appropriate to describe the dark matter distribution (see for example \cite{2017MNRAS.466.1444C}).

Regarding the found parameters, we have obtained similar results to \cite{2011ApJ...729..123Y}, with generally good fittings (inside the 1$\sigma$ values).

A monotonic trend is found between parameter $d$ and radii $r$, indicating stronger correlations inside bigger cells, which contain more structure. For the NBD we have $g = \bar{\xi}_2(V)$, and therefore the opposite trend is found.

For the Log Normal distribution with bias, mean $\bar{N}$ is, as expected, close to the value obtained in Table~\ref{fitcic1} or the best fit of GQED and NBD. Parameter $C_b$, the variance of the matter distribution, strongly varies with $r$. This quantity is roughly related with the cosmological parameter $\sigma_8$, which at $z=0$ is measured $\sigma_8 = 0.828 \pm 0.012$ \citep{2014A&A...571A..16P}. Therefore, for the case of cells of radius $8 h^{-1}$ Mpc, we would expect the value of $C_b$ to be close to $\sigma_8^2 = 0.686$, which we actually found for population 1, with lower redshift, where $C = 0.71 \pm 0.03$. The monotonic evolution of $C_b$ with the cell radius for population 1 shows a good description of the variation of the strength of the clustering with the scale. However, we have not been able to find the same results with population 2. Results on the bias parameter $b$ for different radius of the cells are compatible with expected scale dependent behavior of $b$ \citep{baugh13}.

An interesting comparison can be made between the galaxy variance $C$ and the value $b^2C_b$, the variance of a linearly biased tracer. If we compare these values as found in Table~\ref{fitcic1}, we can see a general similarity of these values except for radii $8$ and $12 h^{-1}$ Mpc of Population 2.

\subsection{Model selection}

The NBD, the GQED and the LND make use of two free parameters, therefore the $\chi^2$ test should be enough to discriminate between them. However, the LN distribution with bias includes a third parameter, and this should be taken into account when comparing its performance with the other distributions. Therefore we use the hypothesis test to discriminate the best fitting distribution for the galaxy Counts-in-Cells distribution. We use the Bayesian and the Akaike Information Criteria ($BIC$ and $AIC$, \cite{schwarz1978,akaike1998information}). The preference on the AIC or the BIC is widely debated \citep{lahiri2001model,konishi2008information}. As usual, we make us of both. These criteria are calculated using the maximum likelihood value $L$, which can be defined with the $\chi^2$ value from the best fit:

\begin{equation}
\log{L} = -\frac{1}{2}\chi^2 + K
\end{equation}
where $K$ is a constant that only depends on the errors of the data values so it cancels out when comparing different models. And the information criteria

\begin{equation}
BIC = -2\log{L} + |\theta|\log{n}
\end{equation}
\begin{equation}
AIC = -2\log{L} + 2|\theta|
\end{equation}

where $L$ is the maximum likelihood for a given sample and model, $|\theta|$ is the number of used parameters in the model and $n$ is the number of observations in the data. Now, we can compare the goodness of fit on any pair of models. We are specially interested in comparing the performance of the three different two-parameters models with the LN distribution with bias (LNb). Then, we will say that a model $\alpha$ is performing a better (or worse) fit than the LN distribution with bias if $BIC_{\alpha} < BIC_{LNb}$ (or $>$), and similarly with $AIC$. It is in these comparisons where constant $K$ cancels. Results are shown in Table~\ref{hypo}.

The values obtained for the GQED and the LN distributions show that the LN distribution with bias is clearly superior, even with a additional parameter, and it is completely justified to use a modified Log Normal. 

However, a comparison between the NBD and LNb distribution shows that the first one is clearly preferable for small radii cells, when the shot noise effect is stronger. For radius $16 h^{-1}$ Mpc and higher, the performance is comparable for both distributions \citep{koopman1943jeffreys,kass1995bayes}. Due to its success among the smaller radii we chose the NBD as the best fit distribution function rather than the LNb distribution, although the satisfactory results at large scales of the later make this distribution still a justifiable election.

\begin{table}
\caption{Model selection} 
\begin{center}
\begin{tabular}{lr|ccc}
\multicolumn{2}{c}{Sample} & \multicolumn{3}{c}{Model}\\
\hline
Pop & $r$ & NBD & GQED & LN\\
\hline
\hline
\multicolumn{5}{c}{$BIC_{\alpha} - BIC_{LNb}$}\\
\hline
1	& 8 & $-118.4234$ & $174.8916$ & $854.8506$\\
	& 12 & $-4.039641$ & $224.7505$ & $440.5025$\\
	& 16 & $-8.94503$ & $89.67867$ & $131.1977$\\
	& 20 & $-0.6724359$ & $48.51936$ & $60.20096$\\
	& 24 & $-4.626123$ & $13.31321$ & $18.69431$\\
2	& 8 & $-17.74021$ & $-1.896006$ & $340.7819$\\
	& 12 & $-21.84225$ & $3.992247$ & $145.9392$\\
	& 16 & $-4.927798$ & $44.1163$ & $122.5448$\\
	& 20 & $3.435232$ & $7.287432$ & $30.46023$\\
	& 24 & $-0.8650677$ & $-1.675998$ & $2.853192$\\

\hline
\multicolumn{5}{c}{$AIC_{\alpha} - AIC_{LNb}$}\\
\hline
1	& 8 & $-115.611$ & $177.704$ & $857.663$\\
	& 12 & $-0.6736$ & $228.1165$ & $443.8685$\\
	& 16 & $-6.1657$ & $92.458$ & $133.977$\\
	& 20 & $1.92298$ & $51.11478$ & $62.79638$\\
	& 24 & $-1.8897$ & $16.04963$ & $21.43073$\\
2	& 8 & $-16.2412$ & $-0.397$ & $342.2809$\\
	& 12 & $-19.7639$ & $6.0706$ & $148.0176$\\
	& 16 & $-2.3736$ & $46.6705$ & $125.099$\\
	& 20 & $6.3106$ & $10.1628$ & $33.3356$\\
	& 24 & $1.58967$ & $0.77874$ & $5.30793$\\
\hline
\end{tabular}
\end{center}
\label{hypo}

{\bf Notes.}
Hypothesis contrast between the NBD, the GQED and the LND with the LN distribution with bias. Values shown correspond to the different between the criterion used for one of the three two-parameters models ($\alpha$) and the LNb distribution. Negative values indicate that the two-parameter model performs a better fit that the LNb distribution, while positive numbers indicate the opposite.
\end{table}

\section{Discussion and conclusions}
\label{sec:conc} 

In this work we proceed with a blind fit of four different probability functions. A blind fit allows us to discriminate the best model of the observed Counts-in-Cells distribution. This work provides a very necessary tool for many different applications in modern cosmology, such as the generation of galaxy mocks or the testing of N-body simulations. 

The GQED and the Log Normal distribution (with no bias parameter) generally show clearly poorer performances and both the $\chi^2$ and the hypothesis test advice against using them as descriptors of the galaxy distribution.

We have also fitted the calculated Counts-in-Cells distribution to the standard Log Normal distribution (equation~\ref{lnpdf}), and the fits are systematically worse than the fits obtained including the bias parameter. We conclude that this is an important result: the NBD provides the best fit for the galaxy Counts-in-Cells distribution for all scales, while the Log Normal distribution modified by the inclusion of the bias term might be a justifiable alternative choice for large scale cells. Moreover, the estimation of parameters $C$ and $b$ from the Log Normal distribution incorporates an interesting asset and allows its direct comparison with other methods of bias estimation \citep{2015A&A...582A..16L,2016ApJ...818..174H}. Nevertheless, we might have obtained results for the bias parameter in tension with other works \citep{2011ApJ...736...59Z}, although our results could be an evidence of the expected scale dependent behavior of the bias parameter \cite{baugh13}.

Regarding the $\chi^2$ results, the NBD is, generally the best model. For the higher radii, $16$ to $24 h^{-1}$ Mpc, the Log Normal distribution with bias performs a slightly better fit. This probability distribution has three free parameters, which could explain its better performance. Actually, our hypothesis test analysis showed that the NBD (two free parameters) is preferable, since the lower $\chi^2$ values of the Log Normal distribution do not justify the extra parameter. Therefore, we find that the NBD is the best model describing the Counts-in-Cells galaxy distribution for all scales, and that a choice for the modified Log Normal distribution is only justified at large scales.

Regarding the physical motivation of the NBD, here we must say that different authors find it fully justified \citep{1992MNRAS.254..247E,betancort2000generalized,1983PhLB..131..116C}, while others have found that this model might contradict known physics \citep{1996ApJ...460...16S}, although these authors also recognize that their calculation of the function $f_V(N)$ for 3D cells performed on the SDSS follows better the NBD than the GQED distribution. The same agreement with the NBD is found for the void-probability function, $f_V(0)$ of the SDSS by \cite{Conroy05}. Moreover, \cite{2016A&A...588A..51B} have recently shown that the Negative Binomial fits rather well the observed Counts-in-Cells distribution function for the VIPERS survey, concluding that the underlying continuous galaxy field should be well described by a Gamma distribution, since the NBD is considered to be the discrete analogue of the continuous Gamma distribution (see \cite{Adell94} where the approximation between these two distributions is studied).

In sum, we conclude that the Negative Binomial Distribution has proved to be a very efficient descriptor of the Counts-in-Cells galaxy distribution.

\begin{acknowledgements}
We would like to thank an anonymous referee for comments and suggestions that have improved the quality and readability of this paper.
This work has been supported by the Spanish Ministerio de Econom\'{\i}a y Competitividad projects AYA2013-48623-C2-2 and AYA2016-81065-C2-2-P, including FEDER contributions, by the Generalitat Valenciana project of excellence PROMETEOII/2014/060.
We acknowledge the use of data from the LasDamas simulations, publicly available at http://lss.phy.vanderbilt.edu/lasdamas/.
We also acknowledge the use of public data from SDSS.
Funding for the SDSS and SDSS-II has been provided by the Alfred P. Sloan Foundation, the Participating Institutions, the National Science Foundation, the U.S. Department of Energy, the National Aeronautics and Space Administration, the Japanese Monbukagakusho, the Max Planck Society, and the Higher Education Funding Council for England. The SDSS Web Site is http://www.sdss.org/.
The SDSS is managed by the Astrophysical Research Consortium for the Participating Institutions. The Participating Institutions are the American Museum of Natural History, Astrophysical Institute Potsdam, University of Basel, University of Cambridge, Case Western Reserve University, University of Chicago, Drexel University, Fermilab, the Institute for Advanced Study, the Japan Participation Group, Johns Hopkins University, the Joint Institute for Nuclear Astrophysics, the Kavli Institute for Particle Astrophysics and Cosmology, the Korean Scientist Group, the Chinese Academy of Sciences (LAMOST), Los Alamos National Laboratory, the Max-Planck-Institute for Astronomy (MPIA), the Max-Planck-Institute for Astrophysics (MPA), New Mexico State University, Ohio State University, University of Pittsburgh, University of Portsmouth, Princeton University, the United States Naval Observatory, and the University of Washington. 
\end{acknowledgements}

\bibliography{ref}

\begin{thebibliography}{60}
\expandafter\ifx\csname natexlab\endcsname\relax\def\natexlab#1{#1}\fi

\bibitem[{{Abazajian} {et~al.}(2009){Abazajian}, {Adelman-McCarthy},
  {Ag{\"u}eros}, {Allam}, {Allende Prieto}, {An}, {Anderson}, {Anderson},
  {Annis}, {Bahcall}, \& et~al.}]{2009ApJS..182..543A}
{Abazajian}, K.~N., {Adelman-McCarthy}, J.~K., {Ag{\"u}eros}, M.~A., {et~al.}
  2009, \apjs, 182, 543

\bibitem[{{Adell} \& {De la Cal}(1994)}]{Adell94}
{Adell}, J.~A. \& {De la Cal}, J. 1994, J. Appl. Prob., 31, 391

\bibitem[{{Ahmad} {et~al.}(2002){Ahmad}, {Saslaw}, \&
  {Bhat}}]{2002ApJ...571..576A}
{Ahmad}, F., {Saslaw}, W.~C., \& {Bhat}, N.~I. 2002, \apj, 571, 576

\bibitem[{Akaike(1998)}]{akaike1998information}
Akaike, H. 1998, in Selected Papers of Hirotugu Akaike (Springer), 199--213

\bibitem[{{Angulo} {et~al.}(2008){Angulo}, {Baugh}, \& {Lacey}}]{Angulo08}
{Angulo}, R.~E., {Baugh}, C.~M., \& {Lacey}, C.~G. 2008, \mnras, 387, 921

\bibitem[{{Arnalte-Mur} {et~al.}(2016){Arnalte-Mur}, {Vielva},
  {Mart{\'{\i}}nez}, {Sanz}, {Saar}, \& {Paredes}}]{2016JCAP...03..005A}
{Arnalte-Mur}, P., {Vielva}, P., {Mart{\'{\i}}nez}, V.~J., {et~al.} 2016,
  \jcap, 3, 005

\bibitem[{{Balian} \& {Schaeffer}(1989)}]{1989A&A...220....1B}
{Balian}, R. \& {Schaeffer}, R. 1989, \aap, 220, 1

\bibitem[{{Baugh}(2013)}]{baugh13}
{Baugh}, C.~M. 2013, \pasa, 30, e030

\bibitem[{{Bel} {et~al.}(2016){Bel}, {Branchini}, {Di Porto}, {Cucciati},
  {Granett}, {Iovino}, {de la Torre}, {Marinoni}, {Guzzo}, {Moscardini},
  {Cappi}, {Abbas}, {Adami}, {Arnouts}, {Bolzonella}, {Bottini}, {Coupon},
  {Davidzon}, {De Lucia}, {Fritz}, {Franzetti}, {Fumana}, {Garilli}, {Ilbert},
  {Krywult}, {Le Brun}, {Le F{\`e}vre}, {Maccagni}, {Ma{\l}ek}, {Marulli},
  {McCracken}, {Paioro}, {Polletta}, {Pollo}, {Schlagenhaufer}, {Scodeggio},
  {Tasca}, {Tojeiro}, {Vergani}, {Zanichelli}, {Burden}, {Marchetti},
  {Mellier}, {Nichol}, {Peacock}, {Percival}, {Phleps}, \&
  {Wolk}}]{2016A&A...588A..51B}
{Bel}, J., {Branchini}, E., {Di Porto}, C., {et~al.} 2016, \aap, 588, A51

\bibitem[{{Bel} \& {Marinoni}(2014)}]{BelMar14}
{Bel}, J. \& {Marinoni}, C. 2014, \aap, 563, A36

\bibitem[{{Bel} {et~al.}(2014){Bel}, {Marinoni}, {Granett}, {Guzzo}, {Peacock},
  {Branchini}, {Cucciati}, {de la Torre}, {Iovino}, {Percival}, {Steigerwald},
  {Abbas}, {Adami}, {Arnouts}, {Bolzonella}, {Bottini}, {Cappi}, {Coupon},
  {Davidzon}, {De Lucia}, {Fritz}, {Franzetti}, {Fumana}, {Garilli}, {Ilbert},
  {Krywult}, {Le Brun}, {Le F{\`e}vre}, {Maccagni}, {Ma{\l}ek}, {Marulli},
  {McCracken}, {Paioro}, {Polletta}, {Pollo}, {Schlagenhaufer}, {Scodeggio},
  {Tasca}, {Tojeiro}, {Vergani}, {Zanichelli}, {Burden}, {Di Porto},
  {Marchetti}, {Mellier}, {Moscardini}, {Nichol}, {Phleps}, {Wolk}, \&
  {Zamorani}}]{Bel14}
{Bel}, J., {Marinoni}, C., {Granett}, B.~R., {et~al.} 2014, \aap, 563, A37

\bibitem[{Betancort-Rijo(2000)}]{betancort2000generalized}
Betancort-Rijo, J. 2000, J. Stat. Phys., 98, 917

\bibitem[{{Blanton} {et~al.}(2005){Blanton}, {Schlegel}, {Strauss},
  {Brinkmann}, {Finkbeiner}, {Fukugita}, {Gunn}, {Hogg}, {Ivezi{\'c}}, {Knapp},
  {Lupton}, {Munn}, {Schneider}, {Tegmark}, \& {Zehavi}}]{2005AJ....129.2562B}
{Blanton}, M.~R., {Schlegel}, D.~J., {Strauss}, M.~A., {et~al.} 2005, \aj, 129,
  2562

\bibitem[{{Carruthers} \& {Duong-van}(1983)}]{1983PhLB..131..116C}
{Carruthers}, P. \& {Duong-van}, M. 1983, Physics Letters B, 131, 116

\bibitem[{{Clerkin} {et~al.}(2017){Clerkin}, {Kirk}, {Manera}, {Lahav},
  {Abdalla}, {Amara}, {Bacon}, {Chang}, {Gazta{\~n}aga}, {Hawken}, {Jain},
  {Joachimi}, {Vikram}, {Abbott}, {Allam}, {Armstrong}, {Benoit-L{\'e}vy},
  {Bernstein}, {Bernstein}, {Bertin}, {Brooks}, {Burke}, {Rosell}, {Carrasco
  Kind}, {Crocce}, {Cunha}, {D'Andrea}, {da Costa}, {Desai}, {Diehl},
  {Dietrich}, {Eifler}, {Evrard}, {Flaugher}, {Fosalba}, {Frieman}, {Gerdes},
  {Gruen}, {Gruendl}, {Gutierrez}, {Honscheid}, {James}, {Kent}, {Kuehn},
  {Kuropatkin}, {Lima}, {Melchior}, {Miquel}, {Nord}, {Plazas}, {Romer},
  {Roodman}, {Sanchez}, {Schubnell}, {Sevilla-Noarbe}, {Smith},
  {Soares-Santos}, {Sobreira}, {Suchyta}, {Swanson}, {Tarle}, \&
  {Walker}}]{2017MNRAS.466.1444C}
{Clerkin}, L., {Kirk}, D., {Manera}, M., {et~al.} 2017, \mnras, 466, 1444

\bibitem[{{Coles} \& {Jones}(1991)}]{1991MNRAS.248....1C}
{Coles}, P. \& {Jones}, B. 1991, \mnras, 248, 1

\bibitem[{{Conroy} {et~al.}(2005){Conroy}, {Coil}, {White}, {Newman}, {Yan},
  {Cooper}, {Gerke}, {Davis}, \& {Koo}}]{Conroy05}
{Conroy}, C., {Coil}, A.~L., {White}, M., {et~al.} 2005, \apj, 635, 990

\bibitem[{{Croton} {et~al.}(2004){Croton}, {Colless}, {Gazta{\~n}aga}, {Baugh},
  {Norberg}, {Baldry}, {Bland-Hawthorn}, {Bridges}, {Cannon}, {Cole},
  {Collins}, {Couch}, {Dalton}, {de Propris}, {Driver}, {Efstathiou}, {Ellis},
  {Frenk}, {Glazebrook}, {Jackson}, {Lahav}, {Lewis}, {Lumsden}, {Maddox},
  {Madgwick}, {Peacock}, {Peterson}, {Sutherland}, \&
  {Taylor}}]{2004MNRAS.352..828C}
{Croton}, D.~J., {Colless}, M., {Gazta{\~n}aga}, E., {et~al.} 2004, \mnras,
  352, 828

\bibitem[{{Dekel} \& {Lahav}(1999)}]{1999ApJ...520...24D}
{Dekel}, A. \& {Lahav}, O. 1999, \apj, 520, 24

\bibitem[{{Di Porto} {et~al.}(2016){Di Porto}, {Branchini}, {Bel}, {Marulli},
  {Bolzonella}, {Cucciati}, {de la Torre}, {Granett}, {Guzzo}, {Marinoni},
  {Moscardini}, {Abbas}, {Adami}, {Arnouts}, {Bottini}, {Cappi}, {Coupon},
  {Davidzon}, {De Lucia}, {Fritz}, {Franzetti}, {Fumana}, {Garilli}, {Ilbert},
  {Iovino}, {Krywult}, {Le Brun}, {Le F{\`e}vre}, {Maccagni}, {Ma{\l}ek},
  {McCracken}, {Paioro}, {Polletta}, {Pollo}, {Scodeggio}, {Tasca}, {Tojeiro},
  {Vergani}, {Zanichelli}, {Burden}, {Marchetti}, {Martizzi}, {Mellier},
  {Nichol}, {Peacock}, {Percival}, {Viel}, {Wolk}, \& {Zamorani}}]{DiPorto16}
{Di Porto}, C., {Branchini}, E., {Bel}, J., {et~al.} 2016, \aap, 594, A62

\bibitem[{Efron \& Tibshirani(1994)}]{efron1994introduction}
Efron, B. \& Tibshirani, R.~J. 1994, An introduction to the bootstrap (CRC
  press)

\bibitem[{{Elizalde} \& {Gaztanaga}(1992)}]{1992MNRAS.254..247E}
{Elizalde}, E. \& {Gaztanaga}, E. 1992, \mnras, 254, 247

\bibitem[{Freedman \& Diaconis(1981)}]{freedman1981histogram}
Freedman, D. \& Diaconis, P. 1981, Probability theory and related fields, 57,
  453

\bibitem[{{Fry} \& {Gaztanaga}(1994)}]{1994ApJ...425....1F}
{Fry}, J.~N. \& {Gaztanaga}, E. 1994, \apj, 425, 1

\bibitem[{{Gazta{\~n}aga} {et~al.}(2002){Gazta{\~n}aga}, {Fosalba}, \&
  {Croft}}]{Gaztanaga02}
{Gazta{\~n}aga}, E., {Fosalba}, P., \& {Croft}, R.~A.~C. 2002, \mnras, 331, 13

\bibitem[{{Hamilton} \& {Tegmark}(2004)}]{2004MNRAS.349..115H}
{Hamilton}, A.~J.~S. \& {Tegmark}, M. 2004, \mnras, 349, 115

\bibitem[{{Hoffmann} {et~al.}(2017){Hoffmann}, {Bel}, \&
  {Gazta{\~n}aga}}]{Hoffmann17}
{Hoffmann}, K., {Bel}, J., \& {Gazta{\~n}aga}, E. 2017, \mnras, 465, 2225

\bibitem[{{Hubble}(1934)}]{1934ApJ....79....8H}
{Hubble}, E. 1934, \apj, 79, 8

\bibitem[{{Hurtado-Gil} {et~al.}(2016){Hurtado-Gil}, {Arnalte-Mur},
  {Mart{\'{\i}}nez}, {Fern{\'a}ndez-Soto}, {Stefanon}, {Ascaso},
  {L{\'o}pez-Sanju{\'a}n}, {M{\'a}rquez}, {Povi{\'c}}, {Viironen}, {Aguerri},
  {Alfaro}, {Aparicio-Villegas}, {Ben{\'{\i}}tez}, {Broadhurst},
  {Cabrera-Ca{\~n}o}, {Castander}, {Cepa}, {Cervi{\~n}o},
  {Crist{\'o}bal-Hornillos}, {Gonz{\'a}lez Delgado}, {Husillos}, {Infante},
  {Masegosa}, {Moles}, {Molino}, {del Olmo}, {Paredes}, {Perea}, {Prada}, \&
  {Quintana}}]{2016ApJ...818..174H}
{Hurtado-Gil}, L., {Arnalte-Mur}, P., {Mart{\'{\i}}nez}, V.~J., {et~al.} 2016,
  \apj, 818, 174

\bibitem[{Kass \& Raftery(1995)}]{kass1995bayes}
Kass, R.~E. \& Raftery, A.~E. 1995, Journal of the american statistical
  association, 90, 773

\bibitem[{{Kayo} {et~al.}(2001){Kayo}, {Taruya}, \&
  {Suto}}]{2001ApJ...561...22K}
{Kayo}, I., {Taruya}, A., \& {Suto}, Y. 2001, \apj, 561, 22

\bibitem[{{Kitaura} {et~al.}(2010){Kitaura}, {Jasche}, \&
  {Metcalf}}]{2010MNRAS.403..589K}
{Kitaura}, F.-S., {Jasche}, J., \& {Metcalf}, R.~B. 2010, \mnras, 403, 589

\bibitem[{Konishi \& Kitagawa(2008)}]{konishi2008information}
Konishi, S. \& Kitagawa, G. 2008, Information criteria and statistical modeling
  (Springer Science \& Business Media)

\bibitem[{Koopman(1943)}]{koopman1943jeffreys}
Koopman, B. 1943, The Journal of Symbolic Logic, 8, 34

\bibitem[{{Kova{\v c}} {et~al.}(2011){Kova{\v c}}, {Porciani}, {Lilly},
  {Marinoni}, {Guzzo}, {Cucciati}, {Zamorani}, {Iovino}, {Oesch}, {Bolzonella},
  {Peng}, {Meneux}, {Zucca}, {Bardelli}, {Carollo}, {Contini}, {Kneib}, {Le
  F{\`e}vre}, {Mainieri}, {Renzini}, {Scodeggio}, {Bongiorno}, {Caputi},
  {Coppa}, {de la Torre}, {de Ravel}, {Finoguenov}, {Franzetti}, {Garilli},
  {Kampczyk}, {Knobel}, {Lamareille}, {Le Borgne}, {Le Brun}, {Maier},
  {Mignoli}, {Pello}, {Perez-Montero}, {Pozzetti}, {Ricciardelli}, {Silverman},
  {Tanaka}, {Tasca}, {Tresse}, {Vergani}, {Abbas}, {Bottini}, {Cappi},
  {Cassata}, {Cimatti}, {Fumana}, {Koekemoer}, {Leauthaud}, {Maccagni},
  {McCracken}, {Memeo}, {Scaramella}, \& {Scoville}}]{2011ApJ...731..102K}
{Kova{\v c}}, K., {Porciani}, C., {Lilly}, S.~J., {et~al.} 2011, \apj, 731, 102

\bibitem[{Lahiri(2001)}]{lahiri2001model}
Lahiri, P. 2001, in Model selection, IMS

\bibitem[{{Layzer}(1956)}]{layzer1956new}
{Layzer}, D. 1956, \aj, 61, 383

\bibitem[{{L{\'o}pez-Sanjuan} {et~al.}(2015){L{\'o}pez-Sanjuan}, {Cenarro},
  {Hern{\'a}ndez-Monteagudo}, {Arnalte-Mur}, {Varela}, {Viironen},
  {Fern{\'a}ndez-Soto}, {Mart{\'{\i}}nez}, {Alfaro}, {Ascaso}, {del Olmo},
  {D{\'{\i}}az-Garc{\'{\i}}a}, {Hurtado-Gil}, {Moles}, {Molino}, {Perea},
  {Povi{\'c}}, {Aguerri}, {Aparicio-Villegas}, {Ben{\'{\i}}tez}, {Broadhurst},
  {Cabrera-Ca{\~n}o}, {Castander}, {Cepa}, {Cervi{\~n}o},
  {Crist{\'o}bal-Hornillos}, {Gonz{\'a}lez Delgado}, {Husillos}, {Infante},
  {M{\'a}rquez}, {Masegosa}, {Prada}, \& {Quintana}}]{2015A&A...582A..16L}
{L{\'o}pez-Sanjuan}, C., {Cenarro}, A.~J., {Hern{\'a}ndez-Monteagudo}, C.,
  {et~al.} 2015, \aap, 582, A16

\bibitem[{{Marinoni} {et~al.}(2005){Marinoni}, {Le F{\`e}vre}, {Meneux},
  {Iovino}, {Pollo}, {Ilbert}, {Zamorani}, {Guzzo}, {Mazure}, {Scaramella},
  {Cappi}, {McCracken}, {Bottini}, {Garilli}, {Le Brun}, {Maccagni}, {Picat},
  {Scodeggio}, {Tresse}, {Vettolani}, {Zanichelli}, {Adami}, {Arnouts},
  {Bardelli}, {Blaizot}, {Bolzonella}, {Charlot}, {Ciliegi}, {Contini},
  {Foucaud}, {Franzetti}, {Gavignaud}, {Marano}, {Mathez}, {Merighi},
  {Paltani}, {Pell{\`o}}, {Pozzetti}, {Radovich}, {Zucca}, {Bondi},
  {Bongiorno}, {Busarello}, {Colombi}, {Cucciati}, {Lamareille}, {Mellier},
  {Merluzzi}, {Ripepi}, \& {Rizzo}}]{2005A&A...442..801M}
{Marinoni}, C., {Le F{\`e}vre}, O., {Meneux}, B., {et~al.} 2005, \aap, 442, 801

\bibitem[{{Markwardt}(2009)}]{mpfit}
{Markwardt}, C.~B. 2009, in Astronomical Society of the Pacific Conference
  Series, Vol. 411, Astronomical Data Analysis Software and Systems XVIII, ed.
  D.~A. {Bohlender}, D.~{Durand}, \& P.~{Dowler}, 251

\bibitem[{{Maurogordato} \& {Lachieze-Rey}(1987)}]{1987ApJ...320...13M}
{Maurogordato}, S. \& {Lachieze-Rey}, M. 1987, \apj, 320, 13

\bibitem[{{McBride} {et~al.}(2011){McBride}, {Berlind}, {Scoccimarro},
  {Manera}, {Tinker}, {Busha}, {Wechsler}, {Wu}, \& {van den
  Bosch}}]{2011AAS...21724907M}
{McBride}, C., {Berlind}, A.~A., {Scoccimarro}, R., {et~al.} 2011, in Bulletin
  of the American Astronomical Society, Vol.~43, American Astronomical Society
  Meeting Abstracts, 249.07

\bibitem[{{Norberg} {et~al.}(2009){Norberg}, {Baugh}, {Gazta{\~n}aga}, \&
  {Croton}}]{2009MNRAS.396...19N}
{Norberg}, P., {Baugh}, C.~M., {Gazta{\~n}aga}, E., \& {Croton}, D.~J. 2009,
  \mnras, 396, 19

\bibitem[{{Peebles}(1980)}]{1980lssu.book.....P}
{Peebles}, P.~J.~E. 1980, {The large-scale structure of the universe}

\bibitem[{{Peyton Jones} {et~al.}(2003)}]{haskell98}
{Peyton Jones}, S. {et~al.} 2003, Journal of Functional Programming, 13, 0,
  \url{http://www.haskell.org/definition/}

\bibitem[{{Planck Collaboration} {et~al.}(2014){Planck Collaboration}, {Ade},
  {Aghanim}, {Armitage-Caplan}, {Arnaud}, {Ashdown}, {Atrio-Barandela},
  {Aumont}, {Baccigalupi}, {Banday}, \& et~al.}]{2014A&A...571A..16P}
{Planck Collaboration}, {Ade}, P.~A.~R., {Aghanim}, N., {et~al.} 2014, \aap,
  571, A16

\bibitem[{Royston(1982)}]{royston1982extension}
Royston, J. 1982, Applied Statistics, 115

\bibitem[{{Saslaw} \& {Fang}(1996)}]{1996ApJ...460...16S}
{Saslaw}, W.~C. \& {Fang}, F. 1996, \apj, 460, 16

\bibitem[{{Saslaw} \& {Hamilton}(1984)}]{1984ApJ...276...13S}
{Saslaw}, W.~C. \& {Hamilton}, A.~J.~S. 1984, \apj, 276, 13

\bibitem[{Schwarz(1978)}]{schwarz1978}
Schwarz, G. 1978, Ann. Statist., 6, 461

\bibitem[{{Sheth}(1995)}]{1995MNRAS.274..213S}
{Sheth}, R.~K. 1995, \mnras, 274, 213

\bibitem[{{Sheth} \& {Saslaw}(1996)}]{1996ApJ...470...78S}
{Sheth}, R.~K. \& {Saslaw}, W.~C. 1996, \apj, 470, 78

\bibitem[{{Sigad} {et~al.}(2000){Sigad}, {Branchini}, \&
  {Dekel}}]{2000ApJ...540...62S}
{Sigad}, Y., {Branchini}, E., \& {Dekel}, A. 2000, \apj, 540, 62

\bibitem[{{Skrutskie} {et~al.}(1997){Skrutskie}, {Schneider}, {Stiening},
  {Strom}, {Weinberg}, {Beichman}, {Chester}, {Cutri}, {Lonsdale}, {Elias},
  {Elston}, {Capps}, {Carpenter}, {Huchra}, {Liebert}, {Monet}, {Price}, \&
  {Seitzer}}]{1997ASSL..210...25S}
{Skrutskie}, M.~F., {Schneider}, S.~E., {Stiening}, R., {et~al.} 1997, in
  Astrophysics and Space Science Library, Vol. 210, The Impact of Large Scale
  Near-IR Sky Surveys, ed. F.~{Garzon}, N.~{Epchtein}, A.~{Omont}, B.~{Burton},
  \& P.~{Persi}, 25

\bibitem[{{Swanson}(Accessed: 2016-03-25)}]{lasdamasweb}
{Swanson}, M. Accessed: 2016-03-25, {LasDamas} webpage,
  \url{http://lss.phy.vanderbilt.edu/lasdamas/}

\bibitem[{{Swanson} {et~al.}(2008){Swanson}, {Tegmark}, {Hamilton}, \&
  {Hill}}]{2008MNRAS.387.1391S}
{Swanson}, M.~E.~C., {Tegmark}, M., {Hamilton}, A.~J.~S., \& {Hill}, J.~C.
  2008, \mnras, 387, 1391

\bibitem[{{White}(1979)}]{1979MNRAS.189..831W}
{White}, S.~D.~M. 1979, \mnras, 189, 831

\bibitem[{{Wild} {et~al.}(2005){Wild}, {Peacock}, {Lahav}, {Conway}, {Maddox},
  {Baldry}, {Baugh}, {Bland-Hawthorn}, {Bridges}, {Cannon}, {Cole}, {Colless},
  {Collins}, {Couch}, {Dalton}, {De Propris}, {Driver}, {Efstathiou}, {Ellis},
  {Frenk}, {Glazebrook}, {Jackson}, {Lewis}, {Lumsden}, {Madgwick}, {Norberg},
  {Peterson}, {Sutherland}, \& {Taylor}}]{2005MNRAS.356..247W}
{Wild}, V., {Peacock}, J.~A., {Lahav}, O., {et~al.} 2005, \mnras, 356, 247

\bibitem[{{Yang} \& {Saslaw}(2011)}]{2011ApJ...729..123Y}
{Yang}, A. \& {Saslaw}, W.~C. 2011, \apj, 729, 123

\bibitem[{{Zehavi} {et~al.}(2011){Zehavi}, {Zheng}, {Weinberg}, {Blanton},
  {Bahcall}, {Berlind}, {Brinkmann}, {Frieman}, {Gunn}, {Lupton}, {Nichol},
  {Percival}, {Schneider}, {Skibba}, {Strauss}, {Tegmark}, \&
  {York}}]{2011ApJ...736...59Z}
{Zehavi}, I., {Zheng}, Z., {Weinberg}, D.~H., {et~al.} 2011, \apj, 736, 59

\end{thebibliography}
\bibliographystyle{aa}

\begin{appendix}
\section{Reliability of the error estimation}\label{app}

The estimation of error bars for our Counts-in-Cells distribution is a task that demands an additional analysis. Two kinds of methods can be followed: intrinsic and extrinsic. In the first group we have the Jackknife resampling method, which together with the Bootstrap method \citep{efron1994introduction}, allows us to estimate the errors in multiple statistics. It can be used when we make use of summary statistics, like the Counts-in-Cells or the correlation function. With these statistics, we perform an analysis over the entire population, shrinking its information into a single quantity, such as the first and second characteristics. These methods are meant to be used when the limited volume of data in use does not allow us to compare our calculations among different populations, and therefore, we do not have a direct way to obtain its variation. As we only have one single galaxy population for each sample, we have to apply an internal error estimate method over the entire population.

On the other hand, the extrinsic methods make use of alternative samples to estimate the variance of our sample of study. In galaxy surveys analysis, these alternative samples can be obtained from mock catalogs, like those presented in section~\ref{data:damas} from LasDamas survey \citep{2011AAS...21724907M}. 

In this section we compare the performance of both methods. To do so, we will compare the error estimations $\sigma_i$ obtained in section~\ref{jack} with those obtained with the Jackknife method over one of the LasDamas simulations.

The used Jackknife method is the `delete one jackknife' method \citep{2009MNRAS.396...19N}. Our data from the chosen LasDamas simulation consists on coordinates in a certain space contained by a window. Dividing this window into different regions we generate $N_{sub}$ subvolumes of different disjoint areas of the sky with the same redshift limits. Each subvolume contains a subsample of cells and deleting one of these subsamples from the parent population we produce a new sample. Cells are included in the corresponding subvolume using their center coordinates, and no further calculation is made regarding their volume overlapping with neighbor subsamples. These resampled data shares its cells and galaxies with the original one but lacks one of the subsamples. We can repeat this procedure $N_{sub}$ times by systematically omitting, in turn, each of the subsamples in which the data has been split. The resampling of the data set consists of $N_{sub} - 1$ remaining subsamples, with volume $\nu(W) - \nu(V_i)$, where $V_i$ is the volume occupied by the $i$th subvolume.

Our parent population of cells $C$ occupies an irregular area of the sky due to the rejection process explained in section~\ref{cic:est}. In order to ensure that all jackknife samples contain the same amount of cells, subsamples are selected in the following way: we sort our cells by declination and create 7 (or other desired number) subgroups of equal amount, in such a way that the groups are also sorted by declination. Then, we perform the same division for each subgroup with the right ascension, sorting each of them and dividing into another 7 subgroups, 49 in total. Now we have subgroups of cells located in differently shaped rectangles but containing the same amount of cells. 
 
Let $f$ be the Counts-in-Cells distribution of our parent population. Once we have generated the new subsamples from the original population, we proceed to calculate the covariance matrix as %

\begin{equation}\label{eq:covmat}
\Sigma_{ij} = \frac{N_{sub}-1}{N_{sub}} \sum_{k=1}^{N_{sub}}(f^k(i) - \bar{f}(i))(f^k(j) - \bar{f}(j))
\end{equation}
where $N_{sub}$ is the number of used subsamples, $f^k$ is the Counts-in-Cells distribution when we omit subsample $k$, and $\bar{f}$ is the mean of distributions $f^k$ for every subsample $N_{sub} - k$. We compute these functions in the values $i$ and $j$ corresponding to all the considered values of $N$, this is, the values of the CiC frequency histogram, from $0$ to the maximum number of galaxies contained into a cell. We finally calculate our standard deviations as $\sigma_{i} = \sqrt{\Sigma_{ii}}$.

\begin{figure}
\begin{center}
\includegraphics[scale=0.7]{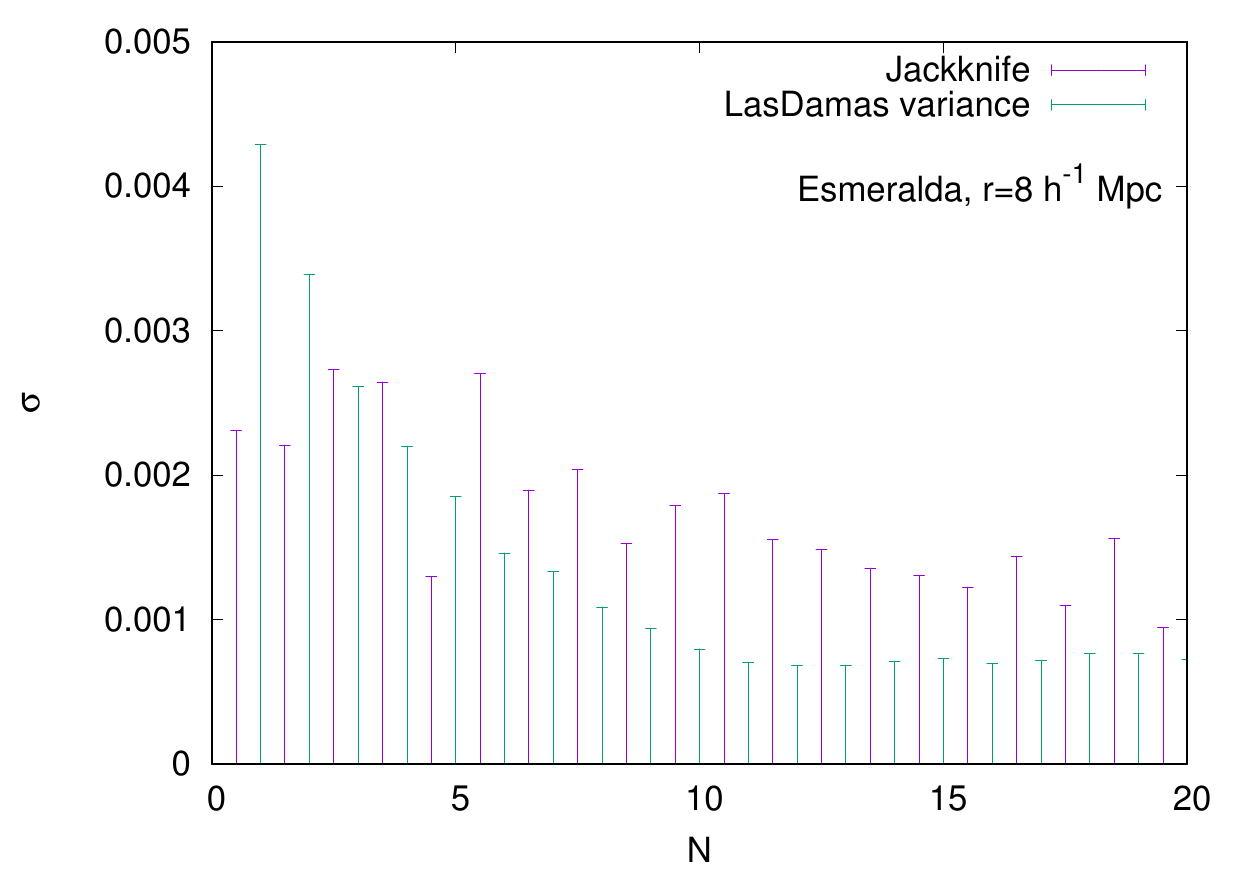}\\
\includegraphics[scale=0.7]{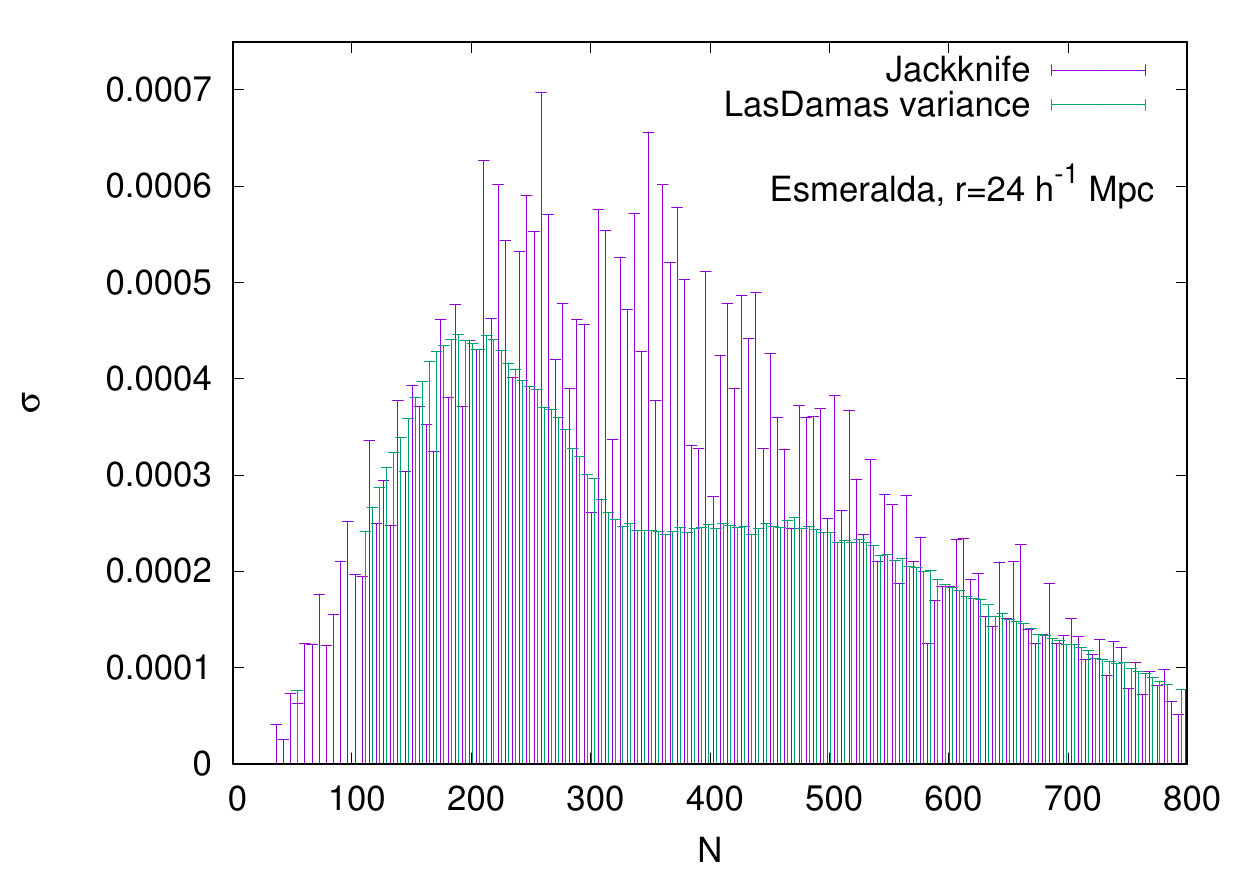}
\end{center}
\caption{\label{errorLD} Comparison between `delete one Jackknife' method error bars for $f_V(N)$ (pink) and its population variance (green) in Esmeralda simulation. Top: $8 h^{-1}$ Mpc cell radius, bottom: $24 h^{-1}$ Mpc cell radius.}

\end{figure}

Now, we can compare the jackknife error estimates with the standard deviations of the LasDamas simulations (see section~\ref{jack}) and check if jackknife errors are over- or underestimated (Fig.~\ref{errorLD}).\\

\begin{figure}
\begin{center}
\includegraphics[scale=0.7]{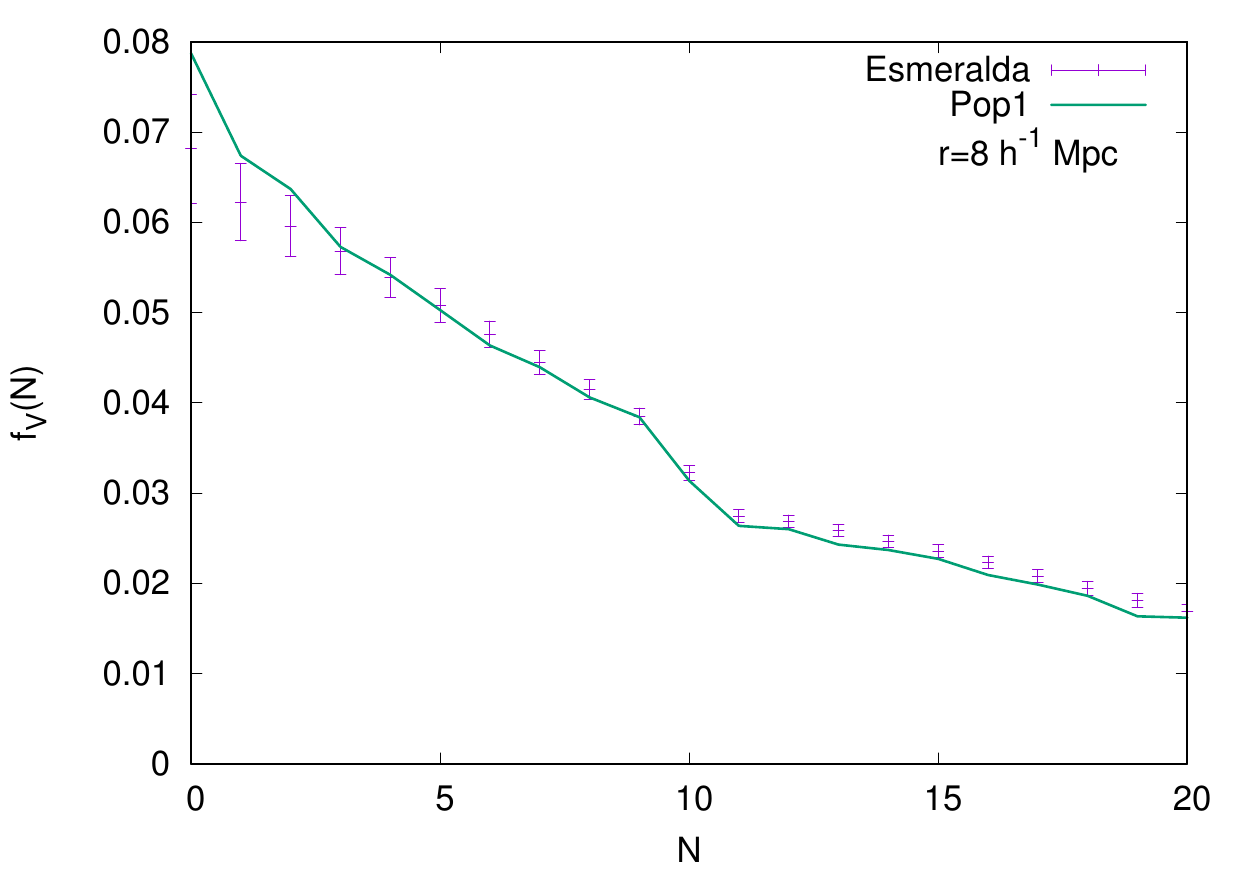}\\
\includegraphics[scale=0.7]{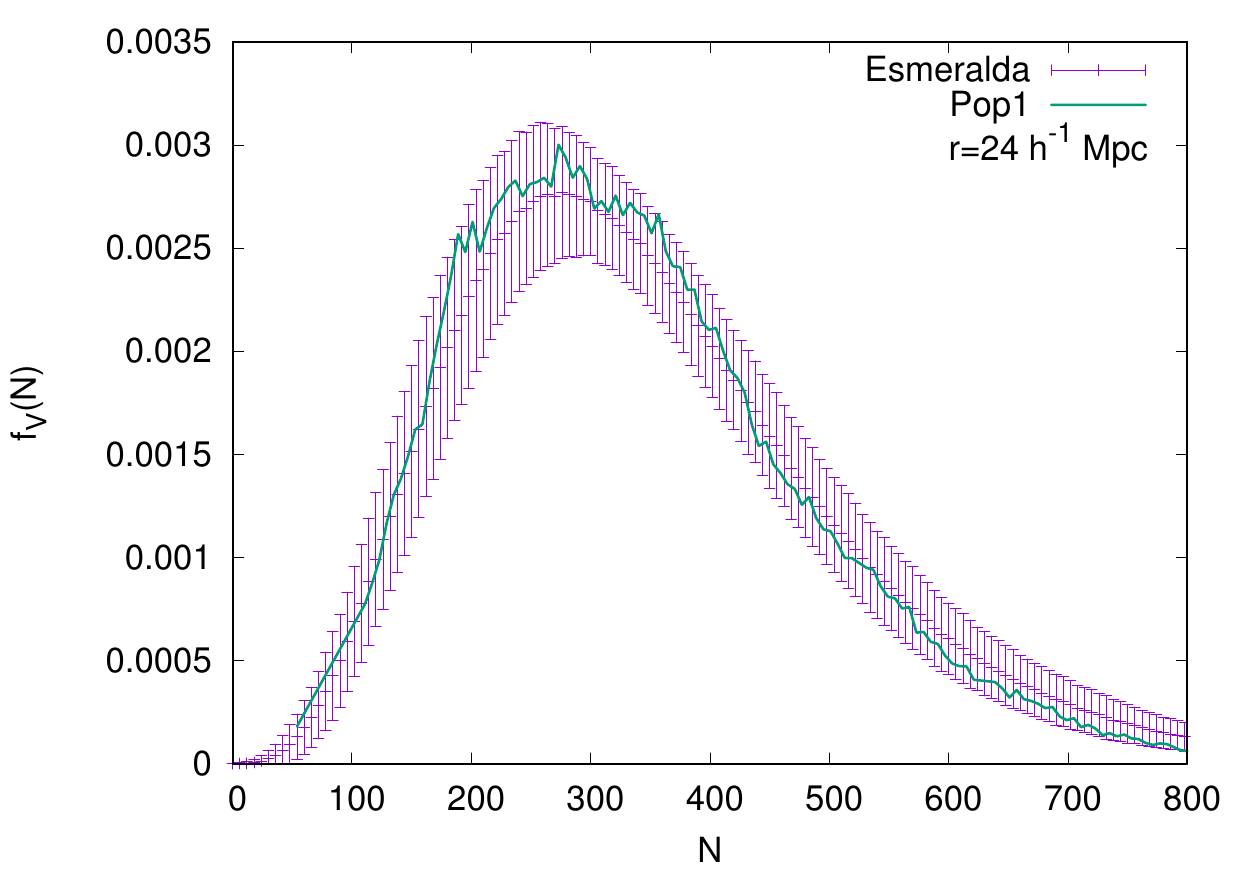}
\end{center}
\caption{\label{compaLD} Comparison between LasDamas CiC distribution (pink) with the results obtained for the SDSS selected samples. Top: $8 h^{-1}$ Mpc cell radius, bottom: $24 h^{-1}$ Mpc cell radius.}
\end{figure}

After these results, we see that the jackknife errors for LasDamas samples are typically larger than the errors obtained from the sample-to-sample variances in the catalogs. This indicates that the same systematic error should be found for the SDSS samples, with jackknife estimated error bars overestimating the real uncertainty. This proves that the jackknife estimations are unreliable in strong mask conditions, even after our cell rejection process. Hence, we will make use of the variances estimated from the LasDamas simulations (equation~\ref{mean}) when fitting our probability functions. In Fig.~\ref{compaLD} we can see the LasDamas CiC distribution compared with the SDSS distribution. The good agreement between data and simulations allows us to use LasDamas variances as error bars for the SDSS Counts-in-Cells distributions.

\end{appendix}

\end{document}